\documentclass[preprint]{emulateapj}
\usepackage{amsmath,natbib,graphicx}
\usepackage{epsf}
\usepackage{color}
\input{epsf}
\usepackage{epsfig}
\usepackage{epstopdf}
\usepackage{lscape}
\DeclareGraphicsExtensions{.jpg,.gif, .pdf,.png,.eps,.ps}
\graphicspath{{FINAL_FIGURES/}}

\newcommand{\mass}{\mbox{$M_{\mbox{\scriptsize 500}}$}}
\newcommand{\sptcl}{SPT\mbox{$_{\mbox{\scriptsize CL}}$}}

\newcommand{\sumnu}{$\Sigma m_{\nu}$}
\newcommand{\fnl}{\mbox{$f_{NL}$}}
\newcommand{\hst}{$H_0$}
\newcommand{\neff}{\ensuremath{N_\mathrm{eff}}}
\newcommand{\LCDM}{\mbox{$\Lambda$CDM}}
\newcommand{\wCDM}{\mbox{$w$CDM}}

\newcommand{\ltsima}{$\; \buildrel < \over \sim \;$}
\newcommand{\ltsim}{\lower.5ex\hbox{\ltsima}}

\newcommand{\sqdeg}{deg$^2$}

\newcommand{\msun}{\ensuremath{M_\odot}}
\newcommand{\SN}{\ensuremath{\xi}}
\newcommand{\nSN}{\ensuremath{\zeta}}
\newcommand{\snm}{$\zeta-M_{500}$}
\newcommand{\Yx}{\mbox{$Y_{X}$}}
\newcommand{\Yxstar}{\mbox{$Y_{X}^*$}}
\newcommand{\yxm}{\mbox{$Y_{X}-M_{500}$}}
\newcommand{\Tx}{\mbox{$T_{X}$}}
\newcommand{\Mg}{\mbox{$M_{g}$}}
\newcommand{\Txtheta}{\mbox{$T_{X}(r)$}}
\newcommand{\Mgr}{\mbox{$M_{g}(r)$}}
\newcommand{\rclust}{\mbox{$r_{\mbox{\tiny 500}}$}}

\newcommand{\chandra}{{\sl Chandra}}
\newcommand{\spitzer}{{\sl Spitzer}}
\newcommand{\hubble}{{\sl Hubble}}
\newcommand{\xmm}{{\sl XMM-Newton}}
\newcommand{\planck}{{\sl Planck}}
\newcommand{\omalllcdm}{$\Omega_m = 0.255 \pm 0.016$}
\newcommand{\sigalllcdm}{$\sigma_8 = 0.795 \pm 0.016$}
\newcommand{\wsptwcdm}{$w = -1.09 \pm 0.36$}
\newcommand{\sigsptwcdm}{$\sigma_8 = 0.773 \pm 0.088$}
\newcommand{\wallwcdm}{$w = -0.973 \pm 0.063$}
\newcommand{\sigallwcdm}{$\sigma_8 = 0.793 \pm 0.028$}

\newcommand{\omallwcdm}{$\Omega_{m} = 0.273 \pm 0.015$}
\newcommand{\hallwcdm}{$h = 0.697 \pm 0.018$}
\newcommand{\fnlspt}{\fnl$=-192 \pm 310$}

\def\KICPChicago{1}
\def\EFIChicago{2}
\def\McGill{3}
\def\Berkeley{4}
\def\UChicago{5}
\def\Munich{6}
\def\MIT{7}
\def\NCSA{8}
\def\CfA{9}
\def\Harvard{10}
\def\ExcellenceCluster{11}
\def\PhysicsUChicago{12}
\def\Miss{13}
\def\AAUChicago{14}
\def\ANL{15}
\def\NIST{16}
\def\PUC{17}
\def\UFlorida{18}
\def\Colorado{19}
\def\NASA{20}
\def\Davis{21}
\def\LBNL{22}
\def\Caltech{23}
\def\Arizona{24}
\def\Michigan{25}
\def\MPE{26}
\def\CaseWestern{27}
\def\Minnesota{28}
\def\STScI{29}
\def\SAIC{30}
\def\Yale{31}

\begin{document}  

\title{Cosmological Constraints from Sunyaev-Zel'dovich-Selected Clusters with X-ray Observations in the 
First 178 Square Degrees of the South Pole Telescope Survey}

\slugcomment{Submitted to \apj}


\author{
B.~A.~Benson\altaffilmark{\KICPChicago,\EFIChicago},
T.~de~Haan\altaffilmark{\McGill},
J.~P.~Dudley\altaffilmark{\McGill},
C.~L.~Reichardt\altaffilmark{\Berkeley},
K.~A.~Aird\altaffilmark{\UChicago},
K.~Andersson\altaffilmark{\Munich,\MIT},
R.~Armstrong\altaffilmark{\NCSA},
M.~L.~N.~Ashby\altaffilmark{\CfA},
M.~Bautz\altaffilmark{\MIT},
M.~Bayliss\altaffilmark{\Harvard}, 
G.~Bazin\altaffilmark{\Munich,\ExcellenceCluster},
L.~E.~Bleem\altaffilmark{\KICPChicago,\PhysicsUChicago},
M.~Brodwin\altaffilmark{\Miss},
J.~E.~Carlstrom\altaffilmark{\KICPChicago,\EFIChicago,\PhysicsUChicago,\AAUChicago,\ANL}, 
C.~L.~Chang\altaffilmark{\KICPChicago,\EFIChicago,\ANL}, 
H.~M. Cho\altaffilmark{\NIST}, 
A.~Clocchiatti\altaffilmark{\PUC},
T.~M.~Crawford\altaffilmark{\KICPChicago,\AAUChicago},
A.~T.~Crites\altaffilmark{\KICPChicago,\AAUChicago},
S.~Desai\altaffilmark{\Munich,\ExcellenceCluster},
M.~A.~Dobbs\altaffilmark{\McGill},
R.~J.~Foley\altaffilmark{\CfA}, 
W.~R.~Forman\altaffilmark{\CfA},
E.~M.~George\altaffilmark{\Berkeley},
M.~D.~Gladders\altaffilmark{\KICPChicago,\AAUChicago},
A.~H.~Gonzalez\altaffilmark{\UFlorida},
N.~W.~Halverson\altaffilmark{\Colorado},
N.~Harrington\altaffilmark{\Berkeley},
F.~W.~High\altaffilmark{\KICPChicago,\AAUChicago}, 
G.~P.~Holder\altaffilmark{\McGill},
W.~L.~Holzapfel\altaffilmark{\Berkeley},
S.~Hoover\altaffilmark{\KICPChicago,\EFIChicago},
J.~D.~Hrubes\altaffilmark{\UChicago},
C.~Jones\altaffilmark{\CfA},
M.~Joy\altaffilmark{\NASA},
R.~Keisler\altaffilmark{\KICPChicago,\PhysicsUChicago},
L.~Knox\altaffilmark{\Davis},
A.~T.~Lee\altaffilmark{\Berkeley,\LBNL},
E.~M.~Leitch\altaffilmark{\KICPChicago,\AAUChicago},
J.~Liu\altaffilmark{\Munich,\ExcellenceCluster},
M.~Lueker\altaffilmark{\Berkeley,\Caltech},
D.~Luong-Van\altaffilmark{\UChicago},
A.~Mantz\altaffilmark{\KICPChicago},
D.~P.~Marrone\altaffilmark{\Arizona},
M.~McDonald\altaffilmark{\MIT},
J.~J.~McMahon\altaffilmark{\KICPChicago,\EFIChicago,\Michigan},
J.~Mehl\altaffilmark{\KICPChicago,\AAUChicago},
S.~S.~Meyer\altaffilmark{\KICPChicago,\EFIChicago,\PhysicsUChicago,\AAUChicago},
L.~Mocanu\altaffilmark{\KICPChicago,\AAUChicago},
J.~J.~Mohr\altaffilmark{\Munich,\ExcellenceCluster,\MPE},
T.~E.~Montroy\altaffilmark{\CaseWestern},
S.~S.~Murray\altaffilmark{\CfA},
T.~Natoli,\altaffilmark{\KICPChicago,\PhysicsUChicago},
S.~Padin\altaffilmark{\KICPChicago,\AAUChicago,\Caltech},
T.~Plagge\altaffilmark{\KICPChicago,\AAUChicago},
C.~Pryke\altaffilmark{\Minnesota}, 
A.~Rest\altaffilmark{\STScI},
J.~Ruel\altaffilmark{\Harvard},
J.~E.~Ruhl\altaffilmark{\CaseWestern}, 
B.~R.~Saliwanchik\altaffilmark{\CaseWestern}, 
A.~Saro\altaffilmark{\Munich},
J.~T.~Sayre\altaffilmark{\CaseWestern}, 
K.~K.~Schaffer\altaffilmark{\KICPChicago,\EFIChicago,\SAIC}, 
L.~Shaw\altaffilmark{\McGill,\Yale},
E.~Shirokoff\altaffilmark{\Berkeley,\Caltech}, 
J.~Song\altaffilmark{\Michigan},
H.~G.~Spieler\altaffilmark{\LBNL},
B.~Stalder\altaffilmark{\CfA},
Z.~Staniszewski\altaffilmark{\CaseWestern},
A.~A.~Stark\altaffilmark{\CfA}, 
K.~Story\altaffilmark{\KICPChicago,\PhysicsUChicago},
C.~W.~Stubbs\altaffilmark{\CfA,\Harvard}, 
R.~Suhada\altaffilmark{\Munich,\ExcellenceCluster},
A.~van~Engelen\altaffilmark{\McGill},
K.~Vanderlinde\altaffilmark{\McGill},
J.~D.~Vieira\altaffilmark{\KICPChicago,\PhysicsUChicago,\Caltech},
A. Vikhlinin\altaffilmark{\CfA},
R.~Williamson\altaffilmark{\KICPChicago,\AAUChicago}, 
O.~Zahn\altaffilmark{\Berkeley},
and
A.~Zenteno\altaffilmark{\Munich,\ExcellenceCluster}
}

\altaffiltext{\KICPChicago}{Kavli Institute for Cosmological Physics, University of Chicago, 5640 South Ellis Avenue, Chicago, IL 60637}
\altaffiltext{\EFIChicago}{Enrico Fermi Institute, University of Chicago, 5640 South Ellis Avenue, Chicago, IL 60637}
\altaffiltext{\McGill}{Department of Physics, McGill University, 3600 Rue University, Montreal, Quebec H3A 2T8, Canada}
\altaffiltext{\Berkeley}{Department of Physics, University of California, Berkeley, CA 94720}
\altaffiltext{\UChicago}{University of Chicago, 5640 South Ellis Avenue, Chicago, IL 60637}
\altaffiltext{\Munich}{Department of Physics,
Ludwig-Maximilians-Universit\"{a}t, Scheinerstr.\ 1, 81679 M\"{u}nchen,
Germany}
\altaffiltext{\MIT}{MIT Kavli Institute for Astrophysics and Space
Research, Massachusetts Institute of Technology, 77 Massachusetts Avenue,
Cambridge, MA 02139}
\altaffiltext{\NCSA}{National Center for Supercomputing Applications,
University of Illinois, 1205 West Clark Street, Urbana, IL 61801}
\altaffiltext{\CfA}{Harvard-Smithsonian Center for Astrophysics, 60 Garden Street, Cambridge, MA 02138}
\altaffiltext{\Harvard}{Department of Physics, Harvard University, 17 Oxford Street, Cambridge, MA 02138}
\altaffiltext{\ExcellenceCluster}{Excellence Cluster Universe,
Boltzmannstr.\ 2, 85748 Garching, Germany}
\altaffiltext{\PhysicsUChicago}{Department of Physics, University of Chicago, 5640 South Ellis Avenue, Chicago, IL 60637}
\altaffiltext{\Miss}{Department of Physics, University of Missouri, 5110 Rockhill Road, Kansas City, MO 64110}
\altaffiltext{\AAUChicago}{Department of Astronomy and Astrophysics, University of Chicago, 5640 South Ellis Avenue, Chicago, IL 60637}
\altaffiltext{\ANL}{Argonne National Laboratory, 9700 S. Cass Avenue, Argonne, IL, USA 60439}
\altaffiltext{\NIST}{NIST Quantum Devices Group, 325 Broadway Mailcode 817.03, Boulder, CO, USA 80305}
\altaffiltext{\PUC}{Departamento de Astronom'a y Astrof'sica, PUC Casilla 306, Santiago 22, Chile}
\altaffiltext{\UFlorida}{Department of Astronomy, University of Florida, Gainesville, FL 32611}
\altaffiltext{\Colorado}{Department of Astrophysical and Planetary Sciences and Department of Physics, University of Colorado, Boulder, CO 80309}
\altaffiltext{\NASA}{Department of Space Science, VP62, NASA Marshall Space Flight Center, Huntsville, AL 35812}
\altaffiltext{\Davis}{Department of Physics, University of California, One Shields Avenue, Davis, CA 95616}
\altaffiltext{\LBNL}{Physics Division, Lawrence Berkeley National Laboratory, Berkeley, CA 94720}
\altaffiltext{\Caltech}{California Institute of Technology, 1200 E. California Blvd., Pasadena, CA 91125}
\altaffiltext{\Arizona}{Steward Observatory, University of Arizona, 933 North Cherry Avenue, Tucson, AZ 85721}
\altaffiltext{\Michigan}{Department of Physics, University of Michigan,
450 Church Street, Ann Arbor, MI, 48109}
\altaffiltext{\MPE}{Max-Planck-Institut f\"{u}r extraterrestrische Physik,
Giessenbachstr.\ 85748 Garching, Germany}
\altaffiltext{\CaseWestern}{Physics Department and CERCA, Case Western Reserve University, 10900 Euclid Ave., Cleveland, OH 44106}
\altaffiltext{\Minnesota}{Physics Department, University of Minnesota, 116 Church Street S.E., Minneapolis, MN 55455}
\altaffiltext{\STScI}{Space Telescope Science Institute, 3700 San Martin
Dr., Baltimore, MD 21218}
\altaffiltext{\SAIC}{Liberal Arts Department, 
School of the Art Institute of Chicago, 
112 S Michigan Ave, Chicago, IL 60603}
\altaffiltext{\Yale}{Department of Physics, Yale University, P.O. Box 208210, New Haven, CT 06520-8120}

\email{bbenson@kicp.uchicago.edu}

\begin{abstract}

We use measurements from the South Pole Telescope (SPT) Sunyaev Zel'dovich (SZ) cluster survey
in combination with X-ray measurements to constrain cosmological parameters. 
We present a statistical method that fits for the scaling relations
of  the SZ and X-ray cluster observables with mass while jointly fitting for cosmology.  The method is 
generalizable to multiple cluster observables, and self-consistently accounts for the effects of the cluster 
selection and uncertainties in cluster mass calibration 
on the derived cosmological constraints.  We apply this method to a data set consisting of an SZ-selected 
catalog of 18 galaxy clusters at $z > 0.3$ from the first 178 \sqdeg\ of the 2500 \sqdeg\ SPT-SZ survey, with 
14 clusters having X-ray observations from either \chandra\ or \xmm.
Assuming a spatially flat \LCDM\ cosmological model, we find the SPT cluster sample constrains 
$\sigma_8 (\Omega_m/0.25)^{0.30} = 0.785 \pm 0.037$.
In combination with measurements of the CMB power 
spectrum from the SPT and the seven-year WMAP data, the SPT cluster sample
constrains \sigalllcdm\ and \omalllcdm, a factor of 1.5 improvement on each parameter over the 
CMB data alone.  We consider several extensions beyond the \LCDM\ model by including the 
following as free parameters: the dark energy equation of state ($w$), the sum of the neutrino masses (\sumnu), the effective number of 
relativistic species (\neff), and a primordial non-Gaussianity ($f_{NL}$). 
We find that adding the SPT cluster data significantly improves the constraints on $w$ and \sumnu\ beyond 
those found when using measurements of the CMB, supernovae, baryon acoustic oscillations, and the Hubble constant.
Considering each extension independently, we best constrain $w=-0.973\pm0.063$ and the sum of 
neutrino masses \sumnu$ < 0.28$ eV at 95\% confidence, a factor of 1.25 and 1.4 improvement, respectively, 
over the constraints without clusters.  Assuming a  \LCDM\ model with a free \neff\ and \sumnu, 
we measure \neff$=3.91 \pm 0.42$ and constrain \sumnu$ < 0.63$ eV at 95\% confidence.
We also use the SPT cluster sample to constrain \fnlspt, consistent with zero primordial non-Gaussianity.   
Finally, we discuss the current systematic limitations due to the cluster mass calibration, and future 
improvements for the recently completed 2500 \sqdeg\ SPT-SZ survey.  The survey 
has detected $\sim$500 clusters with a median redshift of $\sim0.5$ and a median 
mass of $\sim  2.3 \times 10^{14} M_{\odot} / h$ and, when combined with an improved cluster mass 
calibration and existing external cosmological data sets will significantly improve constraints on $w$.  

\end{abstract}


\bigskip\bigskip

\section{Introduction}

\setcounter{footnote}{0}

Clusters of galaxies are the most massive collapsed objects in the universe.  
Their abundance is sensitive to multiple cosmological parameters, in particular 
the matter density, the amplitude of the matter power spectrum, and the dark energy 
equation of state \citep[e.g.,][]{wang98, haiman01, holder01b}. 
Measurements of the cluster abundance that extend to higher redshifts 
become sensitive to dark energy through its effect on the growth of structure.  
This makes cluster abundance measurements an important systematic test 
of the standard dark energy paradigm, because they are affected by dark energy 
in a fundamentally different way than distance-redshift based tests, such as from type Ia supernovae and 
baryon acoustic oscillations.  For the same reason, cluster abundance measurements also 
constrain different cosmological parameter combinations than distance-based tests, and 
their combination can break parameter degeneracies and achieve tighter constraints 
than either method alone \citep[e.g.,][]{linder03}.

Recently there has been significant theoretical and experimental progress in 
efforts to use clusters as cosmological probes.  Large-volume numerical simulations have calibrated 
a ``universal''  cluster mass function over a broad range of cosmologies at 
a level better than current experimental uncertainties \citep[e.g.,][]{jenkins01, warren06, 
tinker08, suman11}.  Numerical simulations have also led to a better understanding 
of systematic biases in cluster mass estimates derived from a broad range of 
cluster observables \citep[e.g.,][]{kravtsov06a, jeltema08, stanek10, becker11}.  
Measurements of the cluster abundance using optical, X-ray, and SZ selection methods
have been used to place competitive constraints on cosmology and dark energy parameters
\citep[e.g.,][]{vikhlinin09, mantz10b, rozo10, vanderlinde10, sehgal10}.  Currently, the most precise
dark energy constraints from clusters are derived from X-ray selected samples which use the
X-ray emission from the hot intra-cluster gas as a tracer of the total mass in the cluster.  
X-ray observables, particularly the gas mass and inferred pressure, 
tend to correlate with cluster mass with low scatter, 
independent of the dynamical state of the cluster or  
the details of non-gravitational physics in clusters \citep[e.g.,][]{kravtsov06a}.  

Hot intra-cluster gas also causes a spectral distortion in the cosmic microwave 
background (CMB) in the direction of clusters from inverse Compton 
scattering, a phenomenon known as the Sunyaev-Zel'dovich (SZ) effect 
\citep{sunyaev72}.  The surface brightness of the SZ effect is 
redshift-independent and largest at mm-wavelengths.  The integrated SZ effect 
from a cluster is effectively measuring the cluster pressure, and is an observable 
that is expected to have comparably low scatter with mass to the 
best X-ray observables \citep{nagai07, shaw08, stanek10}.  Therefore, a mm-wavelength SZ 
survey with sufficient angular resolution is expected to provide clean, mass-limited catalogs out to 
high redshift, probing the regime where the cluster abundance is most sensitive 
to dark energy's effect on the growth rate of structure \citep{carlstrom02}.  

Recently, the first SZ cluster catalogs from three surveys have been released: the South Pole 
Telescope \citep[SPT,][]{staniszewski09, vanderlinde10, williamson11}, the Atacama 
Cosmology Telescope \citep[ACT,][]{marriage11}, and the \planck\ 
satellite \citep{planck11-5.1a}.  However, even with only $\sim$10-20 clusters, 
the cosmological constraints from these surveys have 
been limited by the systematic uncertainty in the cluster mass calibration 
\citep{vanderlinde10, sehgal11}.  X-ray surveys \citep{vikhlinin09, mantz10b} 
have achieved tighter constraints by adopting variations of the following calibration strategy: 
calibrating X-ray observable-mass relations using X-ray hydrostatic mass estimates of relaxed clusters, 
applying this calibration to a larger sample of relaxed and unrelaxed 
clusters, and verifying the overall mass calibration from other methods, 
particularly from weak lensing measurements.  
In this work, we apply a similar strategy to the SPT-SZ survey using the cluster 
sample from \citet{vanderlinde10} (hereafter V10), by incorporating 
an externally calibrated X-ray observable-mass relation and X-ray measurements of the V10 sample in 
order to present improved cosmological constraints.  

This paper is organized as follows.   In Section 2, we describe
the relevant SZ, X-ray, and optical data, analysis methods, and the 
external cosmological data sets used in this work.  
In Section 3, we describe and implement a self-consistent cosmological
analysis using SZ and X-ray observations of the SPT cluster sample
that simultaneously constrains cosmology and the relevant SZ and X-ray 
cluster scaling relations while accounting for the SPT cluster selection function.  
In Section 4, we discuss the constraints on a \LCDM\
cosmological model from the SPT cluster sample, and compare our results to the constraints
from observations of the CMB power spectrum.  In Section 5, we consider 
extensions to the \LCDM\ model by including the following as free parameters:
dark energy equation of state, the sum of the neutrino masses, the effective 
number of relativistic species, and a primordial non-Gaussianity.
We report the relative improvements using the SPT data to constrain each extension.
In Sections 6 and 7, we discuss the limiting systematics and implications for
applying this method to the larger SPT cluster sample.

In this paper, unless otherwise specified, the cluster mass will refer 
to $M_{500}$, the mass enclosed within a spherical radius, $r_{500}$, where the cluster's mean matter density is 
500 times the critical density of the universe at the observed cluster redshift. 
The critical density is $\rho_{crit}(z) = 3H^2(z)/8\pi G$, where
$H(z)$ is the Hubble parameter.

\section{Data and Observations}
\label{sec:data}

\subsection{Cluster Data and Observations}
\label{sec:clust}

The cluster sample used in this work is a sub-sample of a SZ-selected catalog from the 
SPT that was described in V10.  The V10 catalog consisted of 
21 clusters selected by their SZ significance from 178 \sqdeg\ of sky surveyed by the SPT in 2008.
As in V10, we use only the 18 clusters at $z > 0.3$ for the cosmological results in this work.    
The optical and X-ray properties of this catalog have been described previously in \citet{high10} 
and \citet{andersson11}, hereafter H10 and A11,  
respectively.  In this section, we summarize the V10, H10, and A11 
data sets, analysis, and results used in this work.  We also 
report additional spectroscopic redshift and X-ray measurements for several clusters.   

\subsubsection{SZ Observations and the Cluster Sample}
\label{sec:szobs}

The 10-meter diameter SPT is a mm-wavelength telescope designed to conduct a large-area survey 
with low noise and $\sim$1 arcminute angular resolution.  
The SPT receiver consists of a 960 element bolometer array that is sensitive in three bands, at 95, 150, and 220 GHz.  
Details of the telescope and receiver can be found in \citet{padin08}, \citet{carlstrom11}, and \citet{dobbs11}.
The primary goal of the SPT survey is to search for clusters of galaxies via the SZ effect in a 
2500 \sqdeg\ survey that was completed in November 2011.

The first cosmological constraints from the SPT cluster survey were reported in V10, with 
an accompanying cluster catalog.  These results were 
derived from SPT 150 GHz observations of 178 \sqdeg\ observed in 2008, from two approximately equal area fields centered 
at right ascension (R.A.) 5$^h$30$^m$, declination (decl.) -55$^{\circ}$ and R.A. 23$^h$30$^m$, decl. -55$^{\circ}$.
Cluster candidates were identified in the SPT maps by using a matched spatial filter technique 
\citep{haehnelt96,melin06}.  In brief, the SPT maps are filtered in Fourier space to optimize the detection 
of cluster-like objects using a source template constructed from a $\beta$-model of variable 
angular size.  This is done while accounting for the expected signals from the dominant sources of 
astrophysical contamination, instrumental and atmospheric noise, and the effects of the SPT beam 
and timestream filtering.  Candidate galaxy clusters were assigned an SZ significance, \SN, 
defined as the highest signal-to-noise across all filter scales. 

V10 used simulations  to characterize the SZ selection function and the scaling
between \SN\ and cluster mass.  Simulated SZ maps were generated from large-volume dark matter 
simulations \citep{shaw09} using the semi-analytic gas model of \citet{bode07}.
The gas model  was 
calibrated to match the observed X-ray scaling relations for 
low-redshift ($z < 0.25$) clusters.  The cluster selection was characterized by applying the matched filter 
to multiple sky realizations that included the dominant astrophysical components (primary and lensed CMB, thermal SZ, 
and point sources), instrumental and atmospheric noise, and the SPT filtering.  These 
simulations found that at \SN\ $>5$, the SPT catalog was expected to be $\sim95\%$ pure. 
This result is consistent with optical follow-up which found optical cluster counterparts to 21 
of the 22 candidates above this threshold.  The 21 optically confirmed clusters had a median redshift of 
$z=0.74$, and the sample was predicted to be nearly 100\% complete above a mass threshold 
of $M_{500} \sim 6 \times 10^{14} h^{-1}$\msun\ at $z=0.6$.  The simulations 
were also used to put conservative priors on the \SN-mass relation in the V10 cosmological analysis.  
Even with only 18 clusters, the improvement in the cosmological constraints
was limited by the assumed systematic uncertainty on the normalization of the 
\SN-mass relation.

The full cluster catalog used in this work is given in Table \ref{tab:catalog}.  For each 
cluster, we report the name, position, redshift, and the SZ and X-ray observables, where the latter 
assumes a default cosmology.  We note 
that the only SZ product needed for the cosmological analysis described in 
Section \ref{sec:cosmology} is the SZ observable \SN.  In this work 
we improve the cosmological constraints relative to V10 by reducing the 
uncertainty on the \SN-mass relation through inclusion of X-ray observables 
which have an observable-mass relation that has been externally calibrated, 
as described in \citet{vikhlinin09b} and summarized in Section \ref{sec:yxm}.

\begin{table*}
\begin{minipage}{\textwidth}
\centering
\caption{The SPT 178 $\deg^2$ Cluster Catalog and Observables} \small
\begin{tabular}{l cc cc cc}
\hline\hline
\rule[-2mm]{0mm}{6mm}
Object Name  & R.A.    & decl.  & Photo-z         &  Spec-z    &  \SN         & \Yx   \\
&  (deg)   & (deg)  &        &     &          & ($10^{14}$ $M_\sun$ keV)   \\
\hline
SPT CL J0509-5342 &77.336 &-53.705        &  0.47(4) & 0.463 &  6.61 &  $4.3 \pm 0.8$ \\ 
SPT-CL J0511-5154 &77.920 &-51.904        &  0.74(5) & -    &  5.63 & -  \\
SPT-CL J0521-5104 &80.298 &-51.081        &  0.72(5)  & -  &  5.45 & - \\
SPT-CL J0528-5300 &82.017 &-53.000        &  0.75(5)  & 0.765 &  5.45 & $1.6 \pm 0.5$\tablenotemark{b} \\ %
SPT-CL J0533-5005 &83.398 &-50.092        &  0.83(5)  & 0.881 &  5.59 & $1.0 \pm 0.4$\tablenotemark{b}  \\ %
SPT-CL J0539-5744 &85.000 &-57.743        &  0.77(5)  &  -  &  5.12 & - \\
SPT-CL J0546-5345 &86.654 &-53.761        &  1.16(6)  & 1.067\tablenotemark{a} &  7.69 & $4.8 \pm 0.8$\tablenotemark{b}  \\ 
SPT-CL J0551-5709 &87.902 &-57.156        &  0.41(4) & 0.423 &  6.13 & $1.9 \pm 0.4$\tablenotemark{b}  \\ 
SPT-CL J0559-5249 &89.925 &-52.826        &  0.66(4)  & 0.611 &  9.28 & $6.4 \pm 0.8$ \\ 
SPT-CL J2301-5546 &345.469 &-55.776        &  0.78(5) & 0.748\tablenotemark{a} &  5.19 & - \\
SPT-CL J2331-5051 &352.958 &-50.864        &  0.55(4) & 0.571 &  8.04 & $3.5 \pm 0.6$ \\ 
SPT-CL J2332-5358 &353.104 &-53.973        &  0.32(3) & 0.403\tablenotemark{a}  &  7.30 & $6.1 \pm 0.8$\tablenotemark{b}  \\ 
SPT-CL J2337-5942 &354.354 &-59.705        &  0.77(5)  & 0.781  & 14.94 & $8.5 \pm 1.7$ \\ 
SPT-CL J2341-5119 &355.299 &-51.333        &  1.03(5)  & 0.998 &  9.65 & $4.7 \pm 1.0$ \\ 
SPT-CL J2342-5411 &355.690 &-54.189        &  1.08(6) & 1.074\tablenotemark{a} &  6.18 &  $1.4 \pm 0.3$\tablenotemark{b}  \\ 
SPT-CL J2355-5056 &358.955 &-50.937        &  0.35(4)  & 0.320\tablenotemark{a} &  5.89 & $2.2 \pm 0.4$\tablenotemark{b}  \\ 
SPT-CL J2359-5009 &359.921 &-50.160        &  0.76(5)  & 0.774\tablenotemark{a} &  6.35 & $1.8 \pm 0.4$\tablenotemark{b}  \\ 
SPT-CL J0000-5748 &0.250  &-57.807        &  0.74(5) & 0.701\tablenotemark{a}  &  5.48 & $4.2 \pm 1.6$\tablenotemark{b}  \\ 
\hline
\end{tabular}
\label{tab:catalog}
\begin{@twocolumnfalse}
\tablecomments{\SN\ is the maximum signal-to-noise of the SPT-detection obtained over the set of filter scales for each cluster.
The cluster positions in R.A. and decl. are given in degrees and refer 
to the center of the SZ brightness in the SPT map filtered at the preferred scale to maximize the signal-to-noise.
We give the estimated photometric redshift and spectroscopic redshifts, where available.
To be consistent with A11, \Yx\ is calculated assuming a preferred \LCDM\ cosmology 
using  {\sl WMAP}7+BAO+{\sl $H_0$} data with $\Omega_M = 0.272$, $\Omega_\Lambda = 0.728$ and
$H_0 = 70.2~$km$~$s$^{-1}~$Mpc$^{-1}$ \citep{komatsu11}.  
In  \S \ref{sec:lcdm} and \S \ref{sec:xcdm}, \Yx\ is recalculated 
as a function of cosmology and scaling relations for each step in the Markov chain.  
}
\tablenotetext{a}{New spectroscopic redshift since V10.}
\tablenotetext{b}{Updated \Yx\ since A11.}
\end{@twocolumnfalse}
\normalsize
\end{minipage}
\end{table*}

\subsubsection{Optical Redshifts}
\label{sec:oir}

Redshifts of the SPT clusters were measured through a combination of optical photometry and 
spectroscopy.  The majority of the observations and data analyses are identical to those in H10, 
to which we refer the reader for a more detailed description.  
Relative to H10, we include spectroscopic redshift measurements 
for seven additional clusters, which we briefly describe here.  
All cluster redshifts are given in Table \ref{tab:catalog}.  

Optical counterparts and photometric redshifts were measured from a combination of imaging from the Blanco Cosmology 
Survey \citep[BCS, see][]{ngeow09} and targeted observations using the Magellan telescopes.  
Optical images were searched for red sequence galaxies within a $2\arcmin$ radius of the 
SPT candidate location.   A cluster was identified through an excess of red
sequence galaxies relative to the background, and the photometric redshift was estimated by
fitting a red sequence model.  The redshift uncertainty varies over the sample, however it is
typically $\Delta z/(1+z) \sim 0.03$. 

For 15 of the 18 clusters, we have also measured spectroscopic redshifts, which 
we use for the cluster's redshift when measured.
For eight of the clusters we use the spectroscopic redshifts as reported in H10, which were 
measured using the Low Dispersion Survey Spectrograph (LDSS3) on the Magellan Clay 
6.5-m telescope.  For SPT-CL J0546-5345, we use the redshift reported in 
\citet{brodwin10}, measured using multi-slit spectroscopy with the 
Inamori Magellan Areal Camera and Spectrograph (IMACS) on the 
Magellan Baade 6.5-m telescope.  Finally, there are six clusters that have new spectroscopic redshifts, 
which we report in this work for the first time in Table \ref{tab:catalog}.  
These redshifts were measured with a combination of IMACS and GMOS on Gemini South, and the details of the data and analysis 
will be described in \citet[][in prep.]{bazin12}.  

\subsubsection{X-ray Observations}
\label{sec:xray}

X-ray observations were obtained using \chandra\ and \xmm\ for 14 of the clusters 
in Table \ref{tab:catalog}.  The majority of the X-ray observations, data reduction, 
and analyses are the same as described by A11, to which we 
refer the reader for a more detailed description.  Relative to A11, we include 
new \chandra\ observations for five clusters, and re-run the X-ray analysis
for the five clusters with new optical spectroscopic redshifts, one of which also had 
new \chandra\ observations.  
In this section, we summarize the X-ray observations
and results, and describe additional analyses required to incorporate the 
X-ray measurements in the cosmological analysis.  

Summarizing A11, 15 of the 16 highest \SN\ clusters from V10 were targeted for X-ray observations, however, 
in this work, we use only the 14 clusters at $z > 0.3$.  Of these,
twelve were observed with \chandra\, and four clusters were observed with \xmm.  
Two clusters were observed by both \chandra\ and \xmm, and for these clusters only the 
\chandra\ data was included in the analysis. 
From the data, the X-ray observables, \Mg, \Tx, and \Yx, were measured 
in a manner identical to \citet{vikhlinin09b}, where \Mg\ is the gas mass within $r_{500}$, 
\Tx\ is the core-excised X-ray temperature in an annulus between $0.15 - 1.0 \times r_{500}$, 
and \Yx\ $\equiv$ \Mg\Tx.  We solved for each observable and $r_{500}$ iteratively, 
to maintain consistency with their respective observable-mass relations.   Since A11, five of the clusters have 
new spectroscopic redshift measurements.  For these clusters we repeat the A11 reduction 
and analysis using the new redshifts, and give the updated results in Table \ref{tab:xrayobs}.

Five of the clusters from A11 have had additional \chandra\ observations, which we include in this work.  
In Table \ref{tab:xrayids}, we list these 
clusters, the \chandra\ observation IDs, and the improvement in exposure time and cluster source counts 
adding the new observations.  We repeat the A11 reduction and analysis to derive new
constraints on the X-ray observables, which 
are given in Table \ref{tab:xrayobs}.  For these results, relative to A11, we use more 
recent \chandra\ analysis software (CIAO 4.3) and calibration files (CALDB 4.3.3).
We find that the new \chandra\ calibration files typically change \Yx\ by $<5\%$.  This is at a level 
below the assumed mass-normalization uncertainty that we 
assign in Section \ref{sec:yxm}. 

For the cosmological analysis in this work, described in Section \ref{sec:cosmology}, we 
need to calculate the X-ray observables as a function of cosmology 
and scaling relation parameters.  To do this, we derive density and temperature 
profiles for all 14 clusters with X-ray data.  
We calculate \Txtheta\ and \Mgr\ (for the calculation of \Yx$(r)$) from the X-ray 
observations of each cluster assuming a reference cosmology, where $r$ corresponds to 
a physical radius in the cluster and the profiles are defined to return the cluster 
observable within $r$.  
The reference cosmology is chosen to match A11; 
a preferred \LCDM\ cosmology using {\sl WMAP}7+BAO+{\sl $H_0$} data
with $\Omega_M = 0.272$, $\Omega_\Lambda = 0.728$ and
$H_0 = 70.2~$km$~$s$^{-1}~$Mpc$^{-1}$ \citep{komatsu11}.
For three of the clusters with the lowest X-ray photon counts, the \Txtheta\ profiles
have jumps which appear unphysical.
For this reason, we have assumed a functional form of \Txtheta\ that, in combination 
with the measured \Mgr, matches the pressure profile from \citet{arnaud10} 
and is normalized to give the measured \Yx\ assuming the reference cosmology.  
When considering the eleven clusters with well-behaved temperature 
profiles, we find that our cosmological results in Section \ref{sec:lcdm} 
negligibly change when assuming either the functional form of \Txtheta, or the profile derived 
from the data.  Therefore, we consider this approximation valid for this work.  

\subsection{External Cosmological Data Sets}
\label{sec:extcosmo}

In addition to the SPT cluster data set, we incorporate several external 
cosmological data sets, including measurements of the CMB power spectrum (CMB), 
the Hubble constant (\hst), baryon acoustic oscillations (BAO), type Ia supernova (SNe), 
and big bang nucleosynthesis (BBN).  
We will use these abbreviations when referring to these data sets, and will 
use several different combinations of them in our analysis and results.  
Below we give references and a brief description of each external data set.   Also, when 
discussing our results in Section \ref{sec:lcdm} and onward, we will define the \sptcl\ data 
set as the combination of the SPT-SZ data, optical redshift, and X-ray measurements 
described in Section \ref{sec:clust}. 

We use measurements of the CMB power spectrum from the seven-year WMAP data 
release (WMAP7, \citealt{larson11}) and 790 \sqdeg\ of sky observed with the 
SPT \citep{keisler11}.  Following \citet{keisler11},\footnote{http://pole.uchicago.edu/public/data/keisler11} 
we fit the CMB data to a model 
including primary CMB anisotropy plus three nuisance parameters 
that model ``foreground'' signals detectable in the SPT data.
We use low-redshift measurements of $H_0$ from 
the Hubble Space Telescope \citep{riess11}, which we include as a
Gaussian prior of $H_0 = 73.8 \pm 2.4$ km s$^{-1}$ Mpc$^{-1}$.  
We use measurements of the BAO feature using SDSS and 2dFGRS data \citep{percival10}.  
The BAO constraints have been applied as a measurement of $r_s/D_{V}(z=0.2)=0.1905 \pm 0.0061$ 
and $r_s/D_V(z    =0.35)=0.1097 \pm 0.0036$; where $r_s$ 
is the comoving sound horizon size at the baryon drag epoch, $D_V(z) \equiv [(1 + z)^2 D^2_A(z)cz/H(z)]^{1/3}$, 
$D_A(z)$ is the angular diameter distance, and $H(z)$ is the Hubble parameter.  The inverse covariance 
matrix given in Eq.~5 of \citet{percival10} is used for the BAO measurements. 
We use measurements of the luminosity distances of Type Ia supernovae (SNe) from 
the Union2 compilation of 557 SNe \citep{amanullah10}, and include their 
treatment of systematic uncertainties.   Finally, we use a BBN prior from
measurements of the abundances of $^4$He and Deuterium \citep{kirkman03}, 
which we include as a Gaussian prior of $\Omega_b h^2 = 0.022 \pm 0.002$. 

\section{Cosmological Analysis}
\label{sec:cosmology}

In this section, we outline the cosmological analysis method for the SPT 
cluster data set, including the calculation of the cosmological likelihood 
and the assumed parameterization for the cluster mass-observable relations.
This implementation allows for self-consistent constraints on cosmology and the 
cluster scaling relations, i.e., the cluster mass 
calibration, by simultaneously varying the cluster-mass observable relations 
and cosmological parameters using a Markov Chain Monte Carlo (MCMC) technique.
The method is generalizable in way that can include additional cluster observables 
from other data sets (e.g., weak lensing shear, optical velocity dispersions).
We have incorporated our calculation of the SPT cluster likelihood 
into the CosmoMC code\footnote{http://cosmologist.info/cosmomc/} of \citet{lewis02b}
to compute its joint likelihood with the external cosmological data sets.  

\subsection{Scaling Relation Parameterization}
\label{sec:sr}

\subsubsection{X-ray: \yxm}
\label{sec:yxm}

Following \citet{vikhlinin09}, we use \Yx\ as an X-ray proxy for cluster mass, \mass.  We assume a \yxm\ relation of the form
\begin{equation}
\frac{\mass}{ 10^{14} M_{\odot} / h } = \left(A_X h^{3/2}\right) \left( \frac{Y_X}{3 \times 10^{14} M_{\odot}\, {\rm keV}} \right)^{B_{X}} E(z)^{C_{X}},
\label{eqn:yxm}
\end{equation}
parameterized by the normalization $A_{X}$, the slope $B_{X}$, the redshift evolution $C_{X}$, 
where $E(z) \equiv H(z) / H_0$, and a log-normal scatter $D_{X}$ on $Y_X$.
Relative to the form of this equation in 
\citet{vikhlinin09}, we have multiplied the right-hand side by an extra factor of $h$, so that the cluster mass \mass, is in units of 
$M_{\odot}/h$ to match the \snm\ relation in Section \ref{sec:snm}.  For our cosmological analysis, we 
assume Gaussian priors on the scaling relation parameters, which we list in Table \ref{tab:param}.  The priors are motivated by constraints from X-ray measurements by \citet{vikhlinin09b} and simulations, which we describe below. 

\citet{vikhlinin09b} constrained the \yxm\ relation using X-ray observations of a low-redshift ($z < 0.3$) sample of 17 relaxed clusters to estimate the hydrostatic total mass and \Yx.  From simulations, \citet{kravtsov06a} put an upper limit on the systematic offset in the \yxm\ relation between relaxed and unrelaxed clusters of $4\%$.  Simulations also expect that biases in hydrostatic mass estimates are less for relaxed clusters and typically $\lesssim 15\%$ \citep{nagai07}.  Therefore, a \yxm\ relation calibrated from hydrostatic mass estimates of a relaxed cluster sample should have minimal biases and be applicable to a larger cluster sample of both relaxed and unrelaxed clusters.  

From the above measurements, 
\citet{vikhlinin09b} obtained best-fit values of $A_{X} = 5.77 \pm 0.20$ and $B_X=0.57\pm0.03$, where the uncertainties are statistical only.
They estimated the systematic uncertainty in the $A_X$ calibration by comparing to weak lensing mass estimates from \citet{hoekstra07} for a sample of 10 low-redshift clusters.  From this analysis, they estimated a 1$\sigma$ uncertainty of 9\% on the \chandra\ mass scale calibration, which we add in quadrature with their quoted statistical uncertainty on $A_X$.   We have therefore assumed Gaussian priors of  $A_{X} = 5.77 \pm 0.56$ and $B_X=0.57\pm0.03$.

We assume a Gaussian prior of $C_X=-0.4\pm0.2$, consistent with self-similar evolution and a 50\% uncertainty.  
There are relatively few observational constraints on the normalization of the \yxm\ relation at $z>0.3$.  
However, this level of $C_X$ uncertainty was chosen to match the prior assumed in \citet{vikhlinin09}, which was in turn motivated by constraints from the simulations of \citet{kravtsov06a}.  
The 1$\sigma$ prior on $C_X$ would correspond to a 6\% difference in the mass calibration between $z=0.0$ and $0.6$.  For the highest 
redshift cluster in this work, $z=1.074$, this would correspond to an additional 11\% uncertainty in the mass calibration, and a 15\% total uncertainty when also considering the prior on $A_X$.  

We assume a Gaussian prior of $D_X=0.12\pm0.08$, which we truncate below 0.02 and where $D_X=0.12$ corresponds to a 12\% log-normal scatter in \Yx\ for a given mass.  This scatter has been measured to have values ranging from 0-12\% \citep{vikhlinin09b, mantz10}.  Analogous to \citet{vikhlinin09}, for our cosmological analysis we have chosen a prior centered on a value which is consistent with simulations \citep[e.g.,][]{kravtsov06a}.  Although we have assumed a larger uncertainty on the scatter than the range typically found in simulations, this uncertainty is negligible for our cosmological constraints, see Section \ref{sec:err}.  

\subsubsection{SZ: \snm}
\label{sec:snm}
As in V10, we use the detection significance \SN\ as an SZ mass proxy. However, since the relation between \SN\ and 
halo mass is complicated by the comparable effects  of intrinsic scatter and instrumental noise, we introduce the 
unbiased significance, \nSN: the average detection signal-to-noise of a simulated cluster, measured across many 
noise realizations, evaluated at the preferred position and filter scale of that cluster as determined by fitting the cluster in the absence of noise. 

We relate unbiased significance \nSN\ to the detection significance \SN\ as follows. Firstly, \nSN\ is related to $\langle\SN\rangle$ through the relation
\begin{equation}
 \nSN = \sqrt{\langle\SN\rangle^2-3}
\label{eqn:nsn}
\end{equation}
at $\SN>2$. This maximization bias comes from having maximized \SN\ across possible cluster 
positions and filters scales, effectively adding three degrees of freedom to the fit with 
\SN\ analogous to a $\chi^2$.
 Additionally, $\langle\SN\rangle$ relates to \SN\ by a Gaussian scatter of unit width. 
Simulations have been used to verify that these approximations introduce negligible bias or 
scatter compared to the Poisson noise of the sample.  
For further details we refer the reader to V10.

We assume a \snm\ relation of the form
\begin{equation}
\zeta = A_{SZ} \left( \frac{\mass}{3 \times 10^{14} M_{\odot} h^{-1}} \right)^{B_{SZ}} \left(\frac{E(z)}{E(0.6)}\right)^{C_{SZ}},
\label{eqn:zetam}
\end{equation}
parameterized by the normalization $A_{SZ}$, the slope $B_{SZ}$, the redshift evolution $C_{SZ}$, and a log-normal 
scatter, $D_{SZ}$, on $\zeta$.    
V10 motivated 
the form of this relation based on physical arguments, and the expected range of these parameters based on self-similar arguments.   
In V10, the cluster mass was defined within a spherical region in which the density is equal to 200 times the mean matter density at the cluster redshift.
In this work, to be consistent with the \yxm\ relation, we are defining the cluster mass as \mass, the mass in a spherical radius, $r_{500}$, 
within which the density is equal to 500 times the critical 
density of the universe at the cluster redshift.   This change has motivated a change in the redshift 
evolution term from $(1+z)$ to $E(z)$, because of the expected self-similar scaling between $Y_{SZ}$ and 
$M_{500}$ \citep[e.g.,][]{kravtsov06a}.  In addition, we allow for a correlated scatter between \nSN\ and \Yx\ with a correlation 
coefficient $\rho$, which we allow to uniformly vary between 0.02 and 0.98, but away from 0 and 1 for numerical reasons.  

Analogous to V10 and summarized in Section \ref{sec:szobs}, we have used simulated SZ maps to characterize the scaling
between \nSN\ and cluster mass.  
We have repeated this exercise to match the form of the scaling 
given in equation \ref{eqn:zetam}, and we give the Gaussian priors in Table \ref{tab:param}.  
The fractional uncertainty on each parameter matches V10, except for the log-normal scatter, for which 
we allow a larger uncertainty in this work.  However, this uncertainty 
remains negligible for these cosmological constraints, see Section \ref{sec:err}. 

\subsection{Likelihood Model}
\label{sec:like}

The analysis method employed in this work closely mirrors the one presented by V10 with extensions to 
incorporate the X-ray data. In V10, the parameter space was explored through importance sampling of 
pre-existing WMAP MCMC chains.  In this work, we have elected to utilize a full MCMC algorithm. This 
is accomplished through the use of the CosmoMC analysis package, where we have 
included the cluster abundance likelihood as an additional module in the CosmoMC likelihood 
calculation. Among the numerous advantages to this approach is the ability to enforce quantitative 
convergence criteria as well as the optional inclusion of supplemental data sets. 

Each step in the Markov chain selects a new point in the joint cosmological and scaling relation parameter space. Prior to passing these variables to the cluster likelihood evaluation, we use the Code for Anisotropies in the Microwave Background (CAMB) \citep{lewis00} to compute the matter power spectrum at 20 logarithmically spaced redshifts between $0 < z < 2.5$. The matter power spectra, as well as the proposed scaling relation and relevant cosmological parameters, are the inputs to the cluster likelihood function.

At this point, the analysis follows a similar path to that laid out by V10. First, the matter power spectra and cosmology are used to calculate a mass function based upon the \citet{tinker08} prescription, which we calculate for an over-density of $\Delta= 500 \ \Omega_m(z)$, to match our cluster mass definition in Section \ref{sec:sr}. As noted in \citet{tinker08}, this function predicts the halo abundance as a function of input cosmology across a mass range of $10^{11}h^{-1}M_\odot\leq M\leq 10^{15}h^{-1}M_\odot$ and a redshift range of $0\leq z \leq 2.5$.  
\citet{tinker08} claim an overall calibration of their mass function to simulations of $\lesssim5\%$.  \citet{stanek10} found that the inclusion of non-gravitational physics can shift the normalization of the mass function by $\sim10\%$ along the mass direction.  However, this effect is approximately degenerate with an uncertainty between intra-cluster gas observables and mass, which we account for explicitly in our scaling relation uncertainty through equations \ref{eqn:yxm} and \ref{eqn:zetam}.

As in V10, the next step in the analysis is to move the theoretically predicted cluster abundances from their native $M_{500}$ mass space into the observable space for this analysis. V10 define this space by the SZ detection significance, $\SN$, and the optically derived redshift, $z$. This resulted in a two-dimensional surface of predicted cluster abundances in the observable space. In this analysis, we perform a similar transformation, this time  including a third dimension, the X-ray parameter \Yx. This results in a three-dimensional volume of predicted cluster abundances, now as a function of $\SN$, \Yx, and $z$.

Using the scaling relations discussed in \S\ref{sec:sr}, the halo mass function is recast as a predicted number density in terms of \SN, \Yx\ and $z$, which we write as
\begin{equation}
  \frac{dN(\SN,\Yx,z | \vec{p})}{d\SN d\Yx dz} = \int d M P(\SN, \Yx | M, z, \vec{p}) P(M, z | \vec{p}) \Theta(\SN - 5)
  \label{eqn:grid}
\end{equation}
where $\vec{p}$ is the set of cosmological and scaling relation parameters, and $\Theta$ is the Heaviside step function. The likelihood function is then given by the Poisson probability:
\begin{eqnarray}
  \ln \mathcal{L}(\vec{p}) = \sum_i \ln \frac{dN(\SN_i,\Yx_i,z_i, | \vec{p})}{d\SN d\Yx dz} - \nonumber \\
 \int \frac{dN(\SN,\Yx,z, | \vec{p})}{d\SN d\Yx dz} d\SN d\Yx dz,
  \label{eqn:lnlike}
  \end{eqnarray}
where the sum over the $i$ index runs over the SPT cluster catalog. 
Note that we have neglected a global offset to the log-likelihood.

We compute Equation \ref{eqn:grid} on a three-dimensional grid that is 200 by 200 by 30 in the \nSN, \Yx, and $z$ dimensions, respectively. 
For each value of $\Yx$ and $z$ we then convert to the $\SN$ basis by using the $\nSN$-$\SN$ relation defined in Equation \ref{eqn:nsn}, 
where we also convolve with a unit-width Gaussian in \SN\ to account for the noise in the SPT measurement.  

For each step in the MCMC, we recalculate \Yx\ for each cluster given its \Txtheta\ and \Mgr\ profiles 
from Section \ref{sec:xray}, so that its calculated \Yx\ is consistent with the \yxm\ relation and 
$r_{500}$ at that step. 
To account for this in the cosmological likelihood, we modify the likelihood by adding $\sum_i \ln \Yx_i$ to the right hand side 
of Equation \ref{eqn:lnlike}.  For a detailed explanation, see Appendix \ref{app:yx_term}.
For each cluster, we account for finite measurement errors or missing data in $z$ and $\Yx$ by 
modifying the first term in Equation \ref{eqn:lnlike} by marginalizing over 
the relevant parameter, weighted by either a Gaussian likelihood determined 
from its uncertainty or a uniform distribution over the allowed range.

From this calculation we obtain a value for the cluster likelihood corresponding to this particular set of cosmological and 
scaling relation parameters. This value is then returned to CosmoMC where it may be combined with other likelihood 
calculations from supplemental data sets and is used in the MCMC step acceptance/rejection computation. 

\begin{table*}[]
\begin{minipage}{\textwidth}
\centering
\caption{Parameter Table} \small
\begin{tabular}{llll}
\hline\hline
\rule[-2mm]{0mm}{6mm}
Type           & Symbol  &  Meaning  &  Gaussian Prior  \\
\hline
Scaling & $A_{SZ}$ & $\zeta$-mass normalization & $5.58 \pm 1.67$ \\
Relation & $B_{SZ}$ & $\zeta$-mass slope & $1.32 \pm 0.26$ \\
Parameters & $C_{SZ}$ & $\zeta$-mass redshift evolution & $0.87 \pm 0.44$ \\
 & $D_{SZ}$ & Log-normal scatter in $\zeta$ & $0.24 \pm 0.16$ \\
 & $A_{X}$ & $Y_{X}$-mass normalization & $5.77 \pm 0.56$ \\
 & $B_{X}$ & $Y_{X}$-mass slope & $0.57 \pm 0.03$ \\
 & $C_{X}$ & $Y_{X}$-mass redshift evolution & $-0.40 \pm 0.20$ \\
 & $D_{X}$ & Log-normal scatter in $Y_{X}$ & $0.12 \pm 0.08$ \\
 & $\rho$ & Correlated scatter between $\zeta$ and $Y_{X}$ & Uniform:(0.02, 0.98) \\ 
\hline 
Primary & $\Omega_c h^2$ & Dark matter density &  \\
Cosmology & $\Omega_b h^2$ & Baryon density &  \\
Parameters & $100 \Theta_s$ & Angular scale of the sound horizon at last scattering & \\
 & $n_s$ & Scalar tilt of power spectrum &  \\
 & $10^9 \Delta^2_R$ & Scalar amplitude of power spectrum & \\
 & $\tau$ & Optical depth to reionization & \\
\hline
Extension & $w$ & Dark energy equation of state & \\ 
Cosmology & $f_{\nu}$ & Fraction of dark matter in the form of neutrinos, \sumnu$= 94$eV$(f_{\nu} \Omega_c h^2)$ & \\
Parameters & \neff & The effective number of relativistic species & \\
& $f_{NL}$ & Primordial non-Gaussianity parameter & \\
\hline
Derived & $\sigma_8$ & Matter fluctuations on 8 Mpc scales at $z=0$  & \\
Cosmology & $\Omega_m$ & Total matter density & \\
Parameters & $h$ & $h \equiv H_0/100$ km s$^{-1}$ Mpc$^{-1}$, where \hst\ is the Hubble constant at $z=0$ & \\
\hline
\label{tab:param}
\end{tabular}
\end{minipage}
\end{table*}

\section{\LCDM\ Results}
\label{sec:lcdm}

We first consider the \sptcl\ data constraints for a spatially flat \LCDM\ cosmological model.  
For this model, we fit 15 parameters:  the nine scaling relation parameters and six primary cosmology 
parameters listed in Table \ref{tab:lcdm}.   For constraints on any individual parameter, 
we always quote the mean of the likelihood distribution and the 68\% confidence 
interval about the mean.  The confidence interval reflects uncertainties after marginalizing 
over all other parameters, and includes systematic uncertainties in the cluster scaling relations 
and mass calibration, as described in Section \ref{sec:sr}.  In this analysis, we use 
the \sptcl\ and external cosmological data sets as described in Section \ref{sec:data}.  

\begin{table*}[]
\begin{minipage}{\textwidth}
\centering
\caption{\LCDM\ Constraints} \small
\begin{tabular}{lcccc}
\hline\hline
\rule[-2mm]{0mm}{6mm}
Parameter           & Prior   & \sptcl+\hst+BBN & CMB & CMB+\sptcl \\
\hline
$A_{SZ}$  & $5.58 \pm 1.67$ & $5.31\pm0.98$ & - & $4.91\pm0.71$ \\
$B_{SZ}$  & $1.32 \pm0.26$ & $1.39\pm0.15$ & - & $1.40\pm0.15$ \\
$C_{SZ}$  & $0.87 \pm0.44$ & $0.90\pm0.34$ & - & $0.83\pm0.30$ \\
$D_{SZ}$  & $0.24 \pm0.16$ & $0.21\pm0.10$ & - & $0.21\pm0.09$ \\
$A_{X}$  & $5.77 \pm0.56$ & $5.69\pm0.51$ & - & $5.82\pm0.48$ \\
$B_{X}$  & $0.57 \pm0.03$ & $0.564\pm0.029$ & - & $0.563\pm0.029$ \\
$C_{X}$  & $-0.40 \pm0.20$ & $-0.37\pm0.16$ & - & $-0.35\pm0.16$ \\
$D_{X}$  & $0.12 \pm0.08$ & $0.14\pm0.07$ & - & $0.14\pm0.07$ \\
$\rho$  & $(0.02,0.98)$ & $0.52\pm0.27$ & - & $0.52\pm0.27$ \\
\hline
$\Omega_c h^2$ & - & $0.133\pm0.045$ & $0.111\pm0.0048$ & $0.109\pm0.0032$ \\
$\Omega_b h^2$ & - & $0.0221\pm0.0020$ & $0.0222\pm0.0004$ & $0.0223\pm0.0004$ \\
$100 \Theta_s$ & - & $1.065\pm0.041$ & $1.041\pm0.0016$ & $1.041\pm0.0016$ \\
$n_s$ & (0.944, 0.989) & $0.966\pm0.013$ & $0.965\pm0.011$ & $0.967\pm0.010$ \\
$10^9 \Delta^2_R$ & - & $2.16\pm1.30$ & $2.44\pm0.10$ & $2.40\pm0.08$ \\
$\tau$ & - & (0.090) & $0.086\pm0.014$ & $0.087\pm0.014$ \\
\hline
$\sigma_8$ & - & $0.766\pm0.062$ & $0.808\pm0.024$ & $0.795\pm0.016$ \\
$\Omega_m$ & - & $0.285\pm0.083$ & $0.268\pm0.025$ & $0.255\pm0.016$ \\
$h$ & - & $0.739\pm0.024$ & $0.707\pm0.022$ & $0.717\pm0.016$ \\
\hline
\end{tabular}
\label{tab:lcdm}
\tablecomments{The marginalized constraints on the scaling relation and primary
cosmology parameters from Table \ref{tab:param}, where we report the mean of the likelihood distribution 
and the 68\% confidence interval about the mean.  The priors are 
Gaussian, except for $\rho$ and $n_s$, which are uniform over the range given.  
The $n_s$ prior is only used for the \sptcl+\hst+BBN 
data set.  
}
\end{minipage}
\end{table*}

\begin{figure}[]
\centering
\includegraphics[scale=0.47]{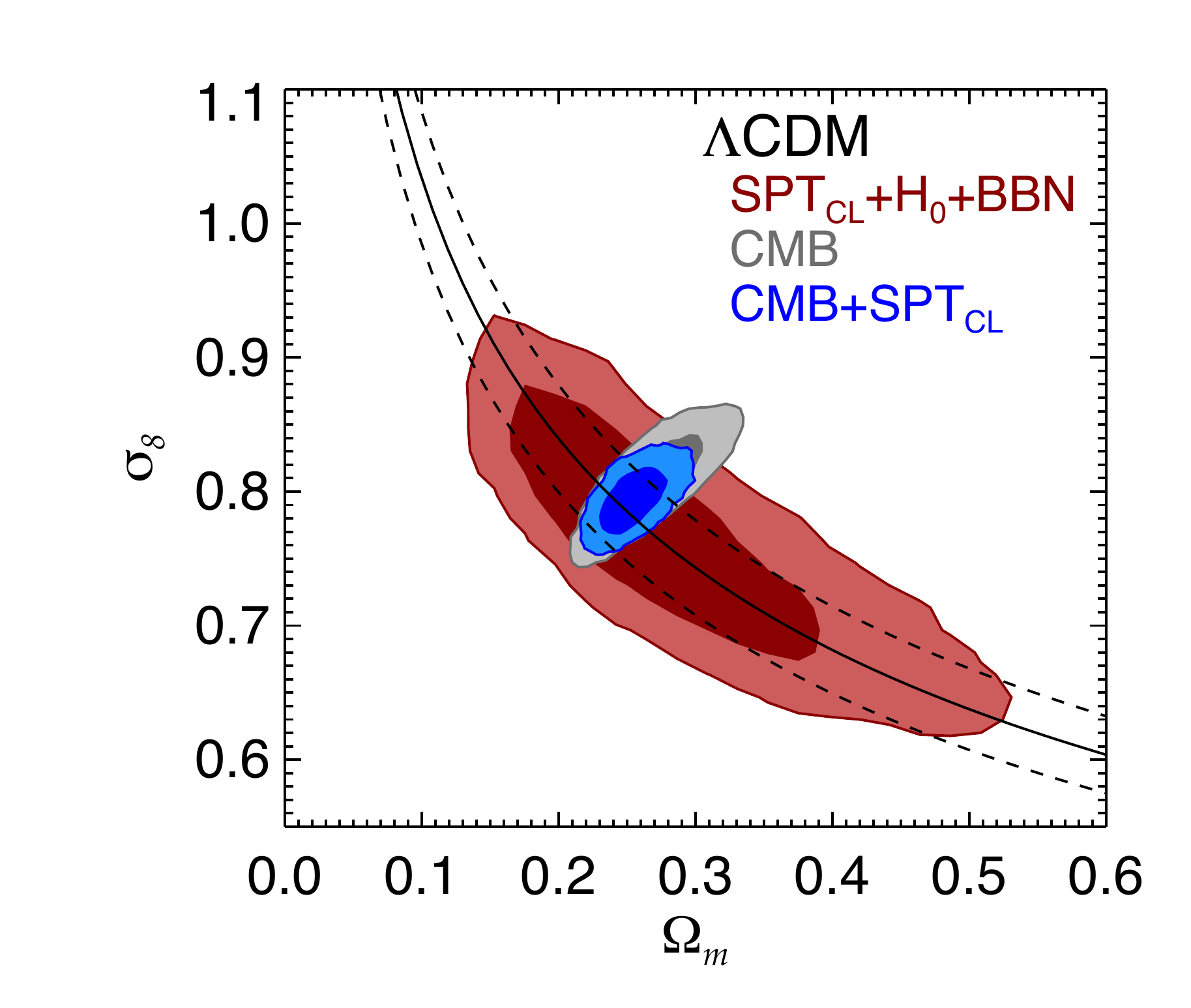}
  \caption[]{Assuming a \LCDM\ cosmology, the two-dimensional marginalized 
  constraints on $\sigma_8$ and $\Omega_{m}$. Contours show the 68\%
  and 95\% confidence regions for the \sptcl+\hst+BBN (red), CMB (gray), and 
  CMB+\sptcl\ (blue) data sets.  The black lines are the best-fit constraint (solid) and 68\% confidence region (dashed) 
  for the combination of parameters that the \sptcl+\hst+BBN data set best 
  constrains: $\sigma_8 (\Omega_m/0.25)^{0.30} = 0.785 \pm 0.037$.
        \\
}
\label{fig:lcdm}
\end{figure}

\subsection{Cosmological Constraints}
\label{sec:lcdmcosmo}

The \sptcl\ data are not sensitive to all six \LCDM\ cosmology parameters.  Here and in Section \ref{sec:wcdm}, 
when considering the \sptcl\ cosmological constraints without CMB data, we always include 
BBN and \hst\ priors, as indicated.  For the \sptcl+\hst+BBN data set, we also fix the optical 
depth of reionization, $\tau$, and allow the scalar tilt, $n_s$, to vary uniformly between 
0.944 and 0.989, the 95\% confidence range from \citet{keisler11} assuming a \LCDM\ model.  However, we note that
the \sptcl\ cosmological constraints vary negligibly over this range of $n_s$.
As noted in Section \ref{sec:extcosmo}, whenever we refer to the \sptcl\ data, we are implicitly 
referring to the combined SPT-SZ data, optical redshift, and X-ray measurements described 
in Section \ref{sec:clust}. 

In Figure \ref{fig:lcdm}, we show the constraints on the $\sigma_8$ and $\Omega_m$ parameters for 
the individual and combined \sptcl\ and CMB data sets.   In Table \ref{tab:lcdm}, we give the marginalized 
constraints for the cosmological and scaling relation parameters.  The latter will be discussed further
in Sections \ref{sec:lcdmsnm} and \ref{sec:err}.  In a \LCDM\ cosmology, the \sptcl\ data is 
most sensitive to $\sigma_8$ and $\Omega_m$.  The number of clusters increases with either parameter, so 
the cluster abundance data effectively constrain a product of the two.  We find that the \sptcl+\hst+BBN constraints
are well approximated as $\sigma_8 (\Omega_m/0.25)^{0.30} = 0.785 \pm 0.037$, 
which we show in Figure \ref{fig:lcdm} by the solid and dashed lines.   Combining the \sptcl\ and CMB data, we 
constrain \sigalllcdm\ and \omalllcdm, a factor of 1.5 improvement on each over the constraints 
from the CMB alone.  

The \sptcl\ constraints are consistent with results
using optical and X-ray selected cluster samples.  Recently, \citet{rozo10} compared 
the cluster constraints from several different methods, and found generally good 
agreement and comparable constraints.  It is typical for cluster based constraints to 
be quoted in terms of the product of $\sigma_8$ and $\Omega_m$ to 
an exponent which varies depending on the mass scale of the cluster sample.  One example
for comparison is \citet{vikhlinin09}, who constrained $\sigma_8 (\Omega_m/0.25)^{0.47} = 0.813 \pm 0.027$.
For typical \LCDM\ model constraints of $\Omega_m \sim 0.25-0.30$, this agrees well with our result.  

\begin{figure}[]
\centering
\includegraphics[scale=0.585]{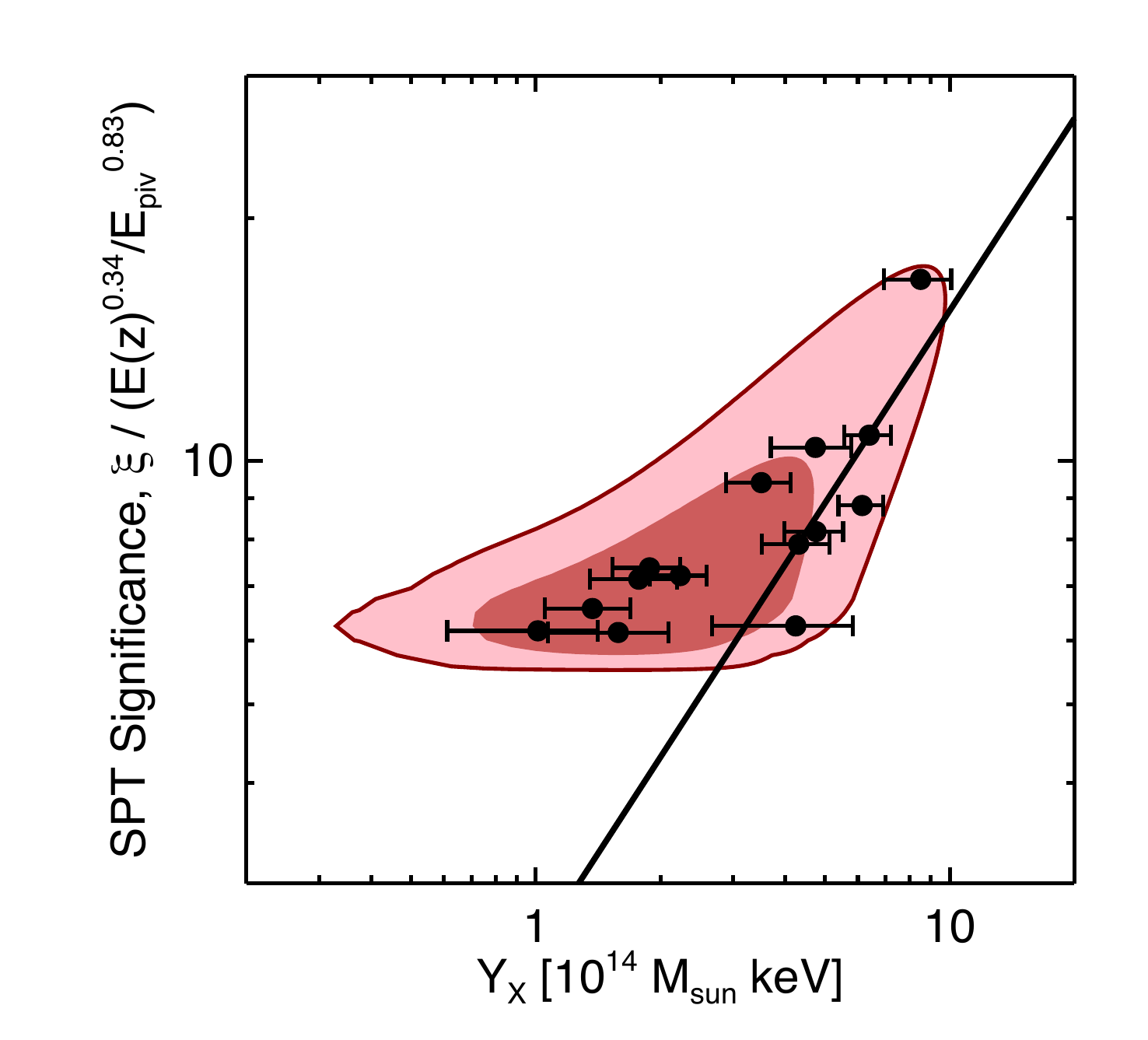}
  \caption[]{A plot of the SZ-significance, \SN, versus the X-ray observable \Yx\ for the 
  14 SPT clusters with X-ray measurements. From the CMB+\sptcl\ data set fit to a \LCDM\ cosmology, 
  we use the best-fit \snm\ and \yxm\ scaling relations to calculate the expected form
  and redshift evolution of the \SN-\Yx relation (solid-line), where $E_{\rm piv} \equiv E(z=0.6)$.
  With the best-fit cosmology parameters, we also  
  predict the effective 68 and 95\%  confidence intervals for the expected distribution of clusters 
  in the \SN-\Yx\ plane (red contours).  
  The measured and predicted cluster distribution show qualitatively good agreement.  
          \\
}
\label{fig:scaling}
\end{figure}

\subsection{Scaling Relation Constraints}
\label{sec:lcdmsnm}

In Figure \ref{fig:scaling}, we show the relationship between \SN\ and \Yx\ for the 14 clusters 
with X-ray observations, over-plotted with the expected distribution of clusters and the best-fit relation
determined using the CMB+\sptcl\ data from Section \ref{sec:lcdmcosmo}.
The combination of the steep mass function and SZ selection yields a distribution of 
clusters visibly offset from the best-fit scaling relation, an effect often referred to 
as Eddington bias.  
We note that our cosmological analysis method described in Section \ref{sec:like}, explicitly accounts 
for the SZ selection and therefore Eddington bias.  We also predict the expected distribution of 
clusters in the \SN-\Yx\ plane assuming the best-fit cosmology and scaling relation 
parameters, and applying a comparable selection as was used for the SPT X-ray 
follow-up ($z > 0.3$ and \SN$>5.45$).   The predicted 14.2 clusters is consistent 
with the 14 detected.  In Figure \ref{fig:scaling}, we over-plot the 
effective 68 and 95\% confidence region in the \SN-\Yx\ plane where we would expect to find these clusters.  
Qualitatively we find good agreement between the observed and predicted cluster distribution.  

In Table \ref{tab:lcdm}, we give the constraints on the \yxm\ and \snm\ scaling relations using
the \sptcl\ and CMB+\sptcl\ data sets.  Because the \yxm\ relation has significantly tighter priors 
than the \snm\ relation, we will not give the \yxm\ constraints for the modified cosmologies 
presented in Section \ref{sec:xcdm}.  
Similarly, for the parameter $\rho$, the correlated scatter 
between \nSN\ and \Yx, we have virtually no constraining power.  In all cases, $\rho$ moves nearly 
uniformly across the entire allowed range, and has a negligible effect 
on the cosmological constraints.  In Appendix \ref{app:mass_ests}, we give posterior 
mass estimates for each cluster using a similar
method as described in V10 and briefly reviewed in the appendix.  

If there were a significant discrepancy 
between the simulation-based prior on the \snm\ relation and the observational prior on the \yxm\ relation, 
we would observe it as an offset between the central value of the \snm\ prior and its best-fit value.  
From the CMB+\sptcl\ constraints, the largest offset is for $A_{SZ}$, with a 
best-fit value of $4.91 \pm 0.71$ compared to the simulation prior of $5.58 \pm 1.67$.  An offset in this direction 
would be consistent with the SZ simulation prior under-estimating the mass of a cluster by a factor of
$\sim((4.91\pm0.71)/5.58)^{1/1.4}=0.91\pm0.09$.  This result is consistent with preliminary estimates from 
A11, who estimated this factor to be $0.78 \pm 0.06$.  We note that A11 did not marginalize 
over uncertainties in either the X-ray scaling relation or cosmological parameters, both of which affect this result.  
The derived offset is also a function of the assumed cosmology.  For example, if we assume a \LCDM\ cosmology 
with a non-zero neutrino mass, as in Section \ref{sec:mnu}, we find a value closer to the simulation prior, $A_{SZ}=5.39 \pm 0.79$, using the 
CMB+H$_0$+\sptcl\ data set.
Therefore, we find no significant inconsistency between the simulation-based prior on the \snm\ relation and the 
observational prior on the \yxm\ relation.

\section{Extensions to \LCDM}
\label{sec:xcdm}

In this section, we consider extensions to a spatially flat \LCDM\ cosmology.  
For each extension, we also fit the nine scaling relation parameters and six primary cosmology 
parameters listed in Table \ref{tab:param}.  We consider four extension cosmologies
where we include the following as free parameters: the dark energy equation of state ($w$), the sum of the 
neutrino masses (\sumnu), the sum of neutrino masses and the effective number 
of relativistic species (\neff), and a primordial non-Gaussianity (\fnl).  
For constraints on any individual parameter, we always quote the mean of the likelihood distribution 
and the 68\% confidence interval about the mean.  The confidence interval will include uncertainties 
after marginalizing over all other parameters, which includes systematic uncertainties in the cluster 
scaling relations and mass calibration, as described in Section \ref{sec:sr}.  In this analysis, we use 
the \sptcl\ and external cosmological data sets as described in Section \ref{sec:data}.  

\begin{table*}[p]
\begin{center}
\caption[\wCDM\ Constraints]{\wCDM\ Constraints} \footnotesize
\begin{tabular}{l c |  c  c | c  c }
\hline \hline
 & & CMB & \hst+BBN & CMB+BAO+SNe & CMB+BAO+SNe \\
 & & & +\sptcl & & +\sptcl \\
 \hline
Scaling & $A_{SZ}$ & - & $5.12\pm1.36$ & - & $4.75\pm0.79$ \\
Parameters & $B_{SZ}$ & - & $1.40\pm0.15$ & - & $1.41\pm0.15$\\
 & $C_{SZ}$ & - & $0.92\pm0.36$ & - & $0.85\pm0.29$ \\
 & $D_{SZ}$ & - & $0.22\pm0.10$ & - & $0.21\pm0.10$ \\
\hline
Cosmology & $\sigma_8$ & $0.864\pm0.120$ &  $0.773\pm0.088$ & $0.823\pm0.040$ & $0.793\pm0.028$ \\
Parameters & $\Omega_m$ &  $0.244\pm0.089$ &  $0.293\pm0.113$ & $0.279\pm0.016$ & $0.273\pm0.015$\\
 & $h$ & $0.775 \pm 0.128$ & $0.740\pm0.025$ & $0.698\pm0.018$ & $0.697\pm0.018$ \\
 & $w$ & $-1.19\pm0.37$ & $-1.09\pm0.36$ &  $-1.014\pm0.078$ & $-0.973\pm0.063$ \\
  \hline \hline
\end{tabular}
\label{tab:wcdm}
\tablecomments{ 
The marginalized constraints on a subset of the scaling relation and cosmology parameters from 
Table \ref{tab:param}.  Scaling relation and primary cosmology parameters not given are still varied in the MCMC and
marginalized over for these constraints.  We report the mean of the likelihood distribution and the 68\% confidence interval about the mean.
}
\end{center}
\end{table*}

\begin{table*}[p]
\begin{center}
\caption[\LCDM\ + \sumnu\ + \neff\ Constraints]{\LCDM\ + \sumnu\ + \neff\ Constraints} \footnotesize
\begin{tabular}{l c | c  c  c | c}
\hline \hline
 &  & CMB+H$_0$+BAO & CMB+H$_0$+BAO & CMB+H$_0$ & CMB+H$_0$+BAO \\
 & &  & +\sptcl & +\sptcl & +\sptcl \\
 \hline
Scaling & $A_{SZ}$ & - & $5.26\pm0.79$& $5.39\pm0.79$ & $5.01\pm0.85$  \\
Parameters & $B_{SZ}$ & - & $1.39\pm0.14$& $1.39\pm0.14$ & $1.41\pm0.15$   \\
 & $C_{SZ}$ & - & $0.89\pm0.30$ & $0.89\pm0.30$  & $0.89\pm0.30$   \\
 & $D_{SZ}$ & - & $0.20\pm0.09$ & $0.20\pm0.10$ & $0.21\pm0.10$  \\
\hline
Cosmology & $\sigma_8$ & $0.761\pm0.043$ & $0.770\pm0.026$ & $0.771\pm0.023$  &  $0.777\pm0.031$  \\
Parameters & $\Omega_m$ & $0.275\pm0.016$ & $0.272\pm0.015$ & $0.260\pm0.018$  &  $0.284\pm0.018$ \\
 & $h$ & $0.698\pm0.014$ & $0.701\pm0.013$& $0.712 \pm 0.017$ & $0.727\pm0.020$  \\
 & \sumnu (eV) & $0.19\pm0.14$ & $0.15\pm0.10$ & $0.12\pm0.09$ & $0.34 \pm 0.17$  \\
 & \sumnu (eV), 95\% CL & $<0.45$ & $<0.33$ & $<0.28$  & $<0.63$\\
 & \neff\ & (3.046) & (3.046) & (3.046) & $3.91 \pm 0.42$ \\
  \hline \hline
\end{tabular}
\label{tab:mnu}
\tablecomments{The marginalized constraints on a subset of the scaling relation and cosmology parameters from 
Table \ref{tab:param}.  Scaling relation and primary cosmology parameters not given are still varied in the MCMC and
marginalized over for these constraints.  We report the mean of the likelihood distribution and the 68\% confidence interval about the mean.}
\end{center}
\end{table*}

\begin{figure}[]
\centering
\includegraphics[scale=0.48]{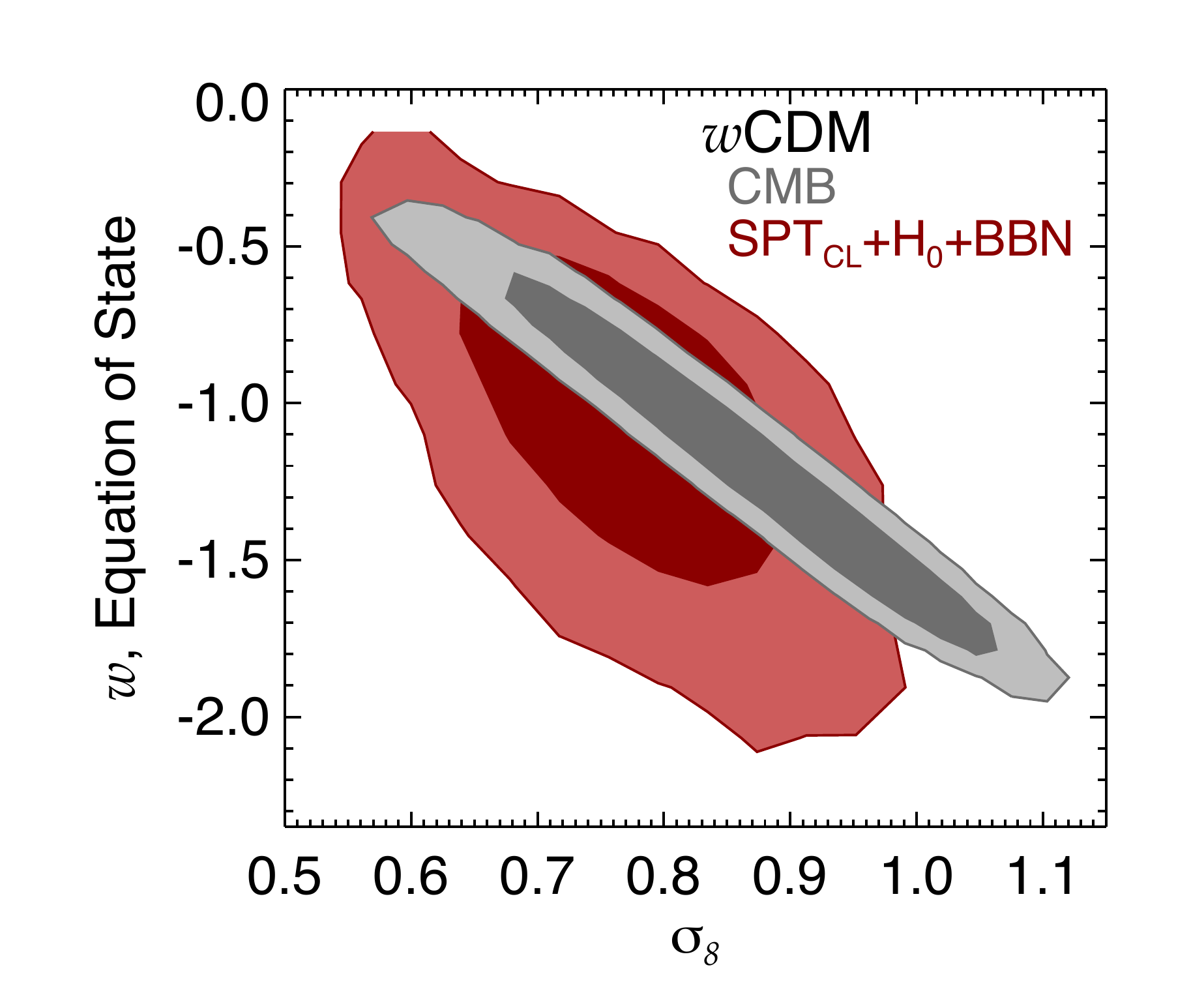}
\caption[]{Assuming a \wCDM\ cosmology, the two-dimensional marginalized 
  constraints on $w$ and $\sigma_8$.  Contours show the 68\%
  and 95\% confidence regions for the \sptcl+\hst+BBN (red) and CMB (gray) data sets.
  \\
  }
 \label{fig:wcdm1}
 \end{figure}
  
\begin{figure*}[ht]
\centering
\includegraphics[scale=0.9]{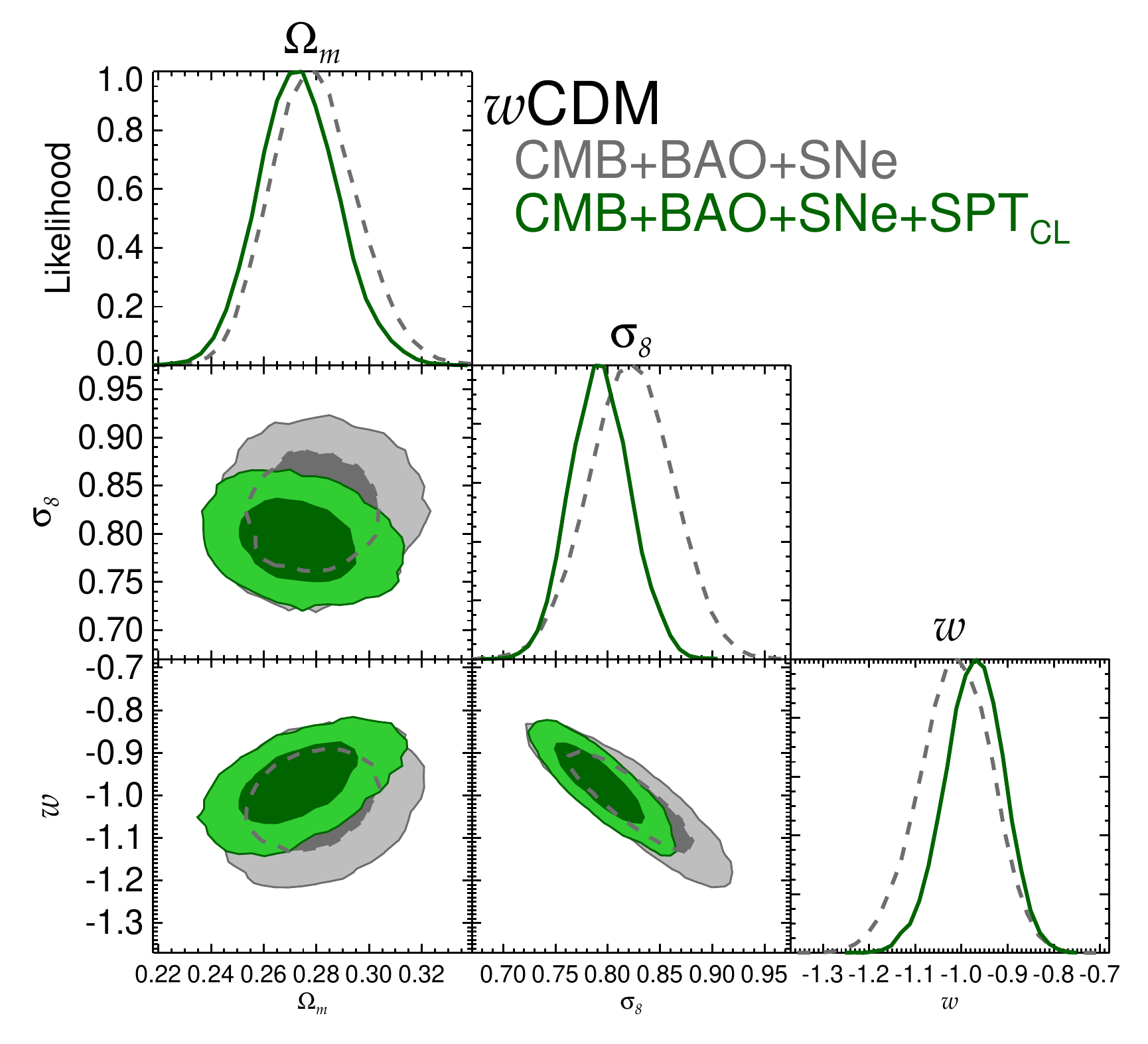}
  \caption[]{Assuming a \wCDM\ cosmology, the constraints on $\Omega_{m}$, $\sigma_8$, 
  and $w$.  The plots along the diagonal are the one-dimensional 
  marginalized likelihood.  The off-diagonal plots are the two-dimensional marginalized constraints showing the 
  68\% and 95\% confidence regions.  We show the constraints for the  
  CMB+BAO+SNe (gray, dashed), and CMB+BAO+SNe+\sptcl\ (green, solid) data sets.
  The \sptcl\ data improves the constraints on $\sigma_8$ and $w$, 
  by factors of 1.4 and 1.25, respectively.
        \\
}
\label{fig:wcdm2}
\end{figure*}

\begin{figure}[]
\centering
\includegraphics[scale=0.48]{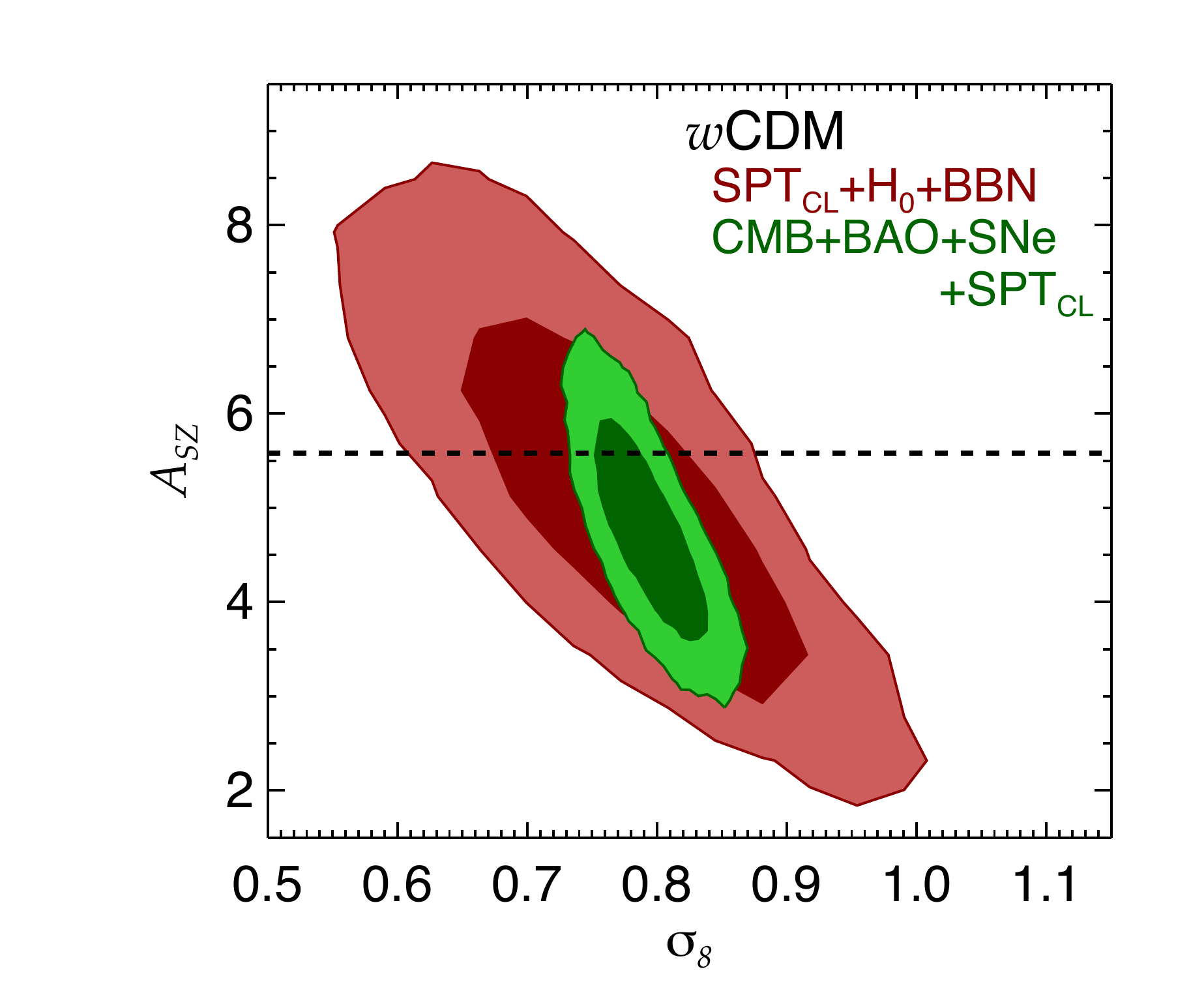}
\caption[]{Assuming a \wCDM\ cosmology, the two-dimensional marginalized 
  constraints on $A_{SZ}$ and $\sigma_8$.  Contours show the 68\%
 and 95\% confidence regions for the \sptcl+\hst+BBN (red) and CMB+BAO+SNe+\sptcl\ (green) data sets.  The 
 horizontal black dashed line is the center of the theory prior on $A_{SZ}$.  
        \\
}
\label{fig:wcdm1b}
\end{figure}

\subsection{\wCDM}
\label{sec:wcdm}

The first extension we consider is a \wCDM\ cosmology, a model in which the 
equation of state of dark energy is a constant $w$.  
The cluster abundance and the shape of the mass function depend on $w$ 
through its effect on the growth of structure, or equivalently 
the redshift evolution of $\sigma_8$.  The CMB data measure structure at $z \sim 1100$, 
and therefore require significant extrapolation to predict the cluster abundance 
in the redshift range of the SPT sample ($0.3 < z < 1.1$).  Therefore, consistency between the 
implied $w$ from both data sets is already an important systematic test 
of dark energy.  

In Figure \ref{fig:wcdm1}, we show the constraints on $w$ and $\sigma_8$ 
using the CMB and \sptcl+\hst+BBN data sets.  The likelihood contours 
have significant overlap, implying the data are in good agreement.  Relative 
to the CMB, the \sptcl\ data tend to disfavor cosmologies with large $\sigma_8$ and more negative $w$.
In Table \ref{tab:wcdm}, we give 
marginalized constraints for several cosmological and scaling relation parameters.  The
\sptcl\ data constrain \wsptwcdm, and have similar constraining power to the CMB data, 
for which the constraints have a significant degeneracy between $w$ and $\sigma_8$.  
The \sptcl\ data simultaneously constrain \sigsptwcdm.  This constraint has a factor of $\sim$1.4 lower uncertainty
than that from the CMB data.

\subsubsection{\wCDM\ with BAO and SNe data sets}
\label{sec:wcdmsne}

In this section, we consider the improvement in \wCDM\ cosmological constraints when adding 
the \sptcl\ data to the CMB, BAO, and SNe data sets.  In Figure \ref{fig:wcdm2}, we show the 
constraints of the combined CMB+BAO+SNe data set, before and after including the \sptcl\ data. 
The \sptcl\ data most significantly improve the constraints on $\sigma_8$ and $w$; 
reducing the allowed two-dimensional likelihood area by a factor of $\sim$1.8.   In Table \ref{tab:wcdm}, we give 
the marginalized constraints for several parameters before and after the inclusion of the \sptcl\ data. 
The combined constraints are \wallwcdm\ and \sigallwcdm, a factor of 1.25 and 1.4 improvement, respectively, 
over the constraints without clusters.  The combined data set also constrains 
\omallwcdm\ and \hallwcdm.
These constraints are consistent with previous cluster-based results \citep{vikhlinin09, mantz10b, rozo10}, 
which used X-ray and optically selected samples of typically lower redshift clusters. 
The sensitivity of the \sptcl\ cluster data to the amplitude of structure, $\sigma_8$, 
is primarily what gives it the ability to break degeneracies with the distance-relation based 
constraints from the BAO and SNe data sets.   
We note the slight tension with the \hst\ constraints from \citet{riess11} of  $h = 0.738 \pm 0.024$.  
While this tension is not significant, it helps to intuitively explain some constraints on 
neutrino mass in Section \ref{sec:mnu}.  

\subsubsection{\snm\ Constraints}
\label{sec:wcdmsnm}

Given the work of V10 and other cluster results \citep[e.g.,][]{vikhlinin09, mantz10b, rozo10}, we 
expect the cluster mass-calibration to be the dominant systematic 
uncertainty limiting our results.  In Figure \ref{fig:wcdm1b}, we 
show the constraints on $A_{SZ}$ and $\sigma_8$.  The \sptcl+\hst+BBN data set has a
significant degeneracy between its constraints on $A_{SZ}$ and $\sigma_8$.  
From this data, we constrain the fractional uncertainty, $\delta A_{SZ} / A_{SZ}$, to be 27\%, which is 
effectively constrained only by the 14 clusters that have both X-ray and SZ measurements.  
This constraint is not significantly better than the uncertainty in the simulation based prior of 30\%.  
With enough X-ray observations, we expect the \snm\ calibration to be limited by the uncertainty 
of the \yxm\ relation, 
because the latter is currently better observationally constrained.  In this limit, we would expect 
a fractional uncertainty on $A_{SZ}$ of $B_{SZ} \times (\delta A_{X} / A_{X}) \sim 14\%$. 
The above would suggest that 
for a \wCDM\ cosmology we would need X-ray observations of $\sim50$ clusters, 
i.e., $\sim14$ clusters $\times (27\%/14\%)^2$, to calibrate $A_{SZ}$ in terms of
mass so that its not the dominant source of uncertainty.  When adding the BAO and SNe data 
sets, we improve the constraints on $A_{SZ}$ to an accuracy 
of $\sim16\%$.  
However, these data sets are not sensitive to either $A_{SZ}$ or $\sigma_8$, and cannot completely break 
their degeneracy.  In Section \ref{sec:err}, we will discuss the systematic uncertainties 
 from this degeneracy on our cosmological constraints in more detail.  

\begin{figure}[]
\centering
\includegraphics[scale=0.48]{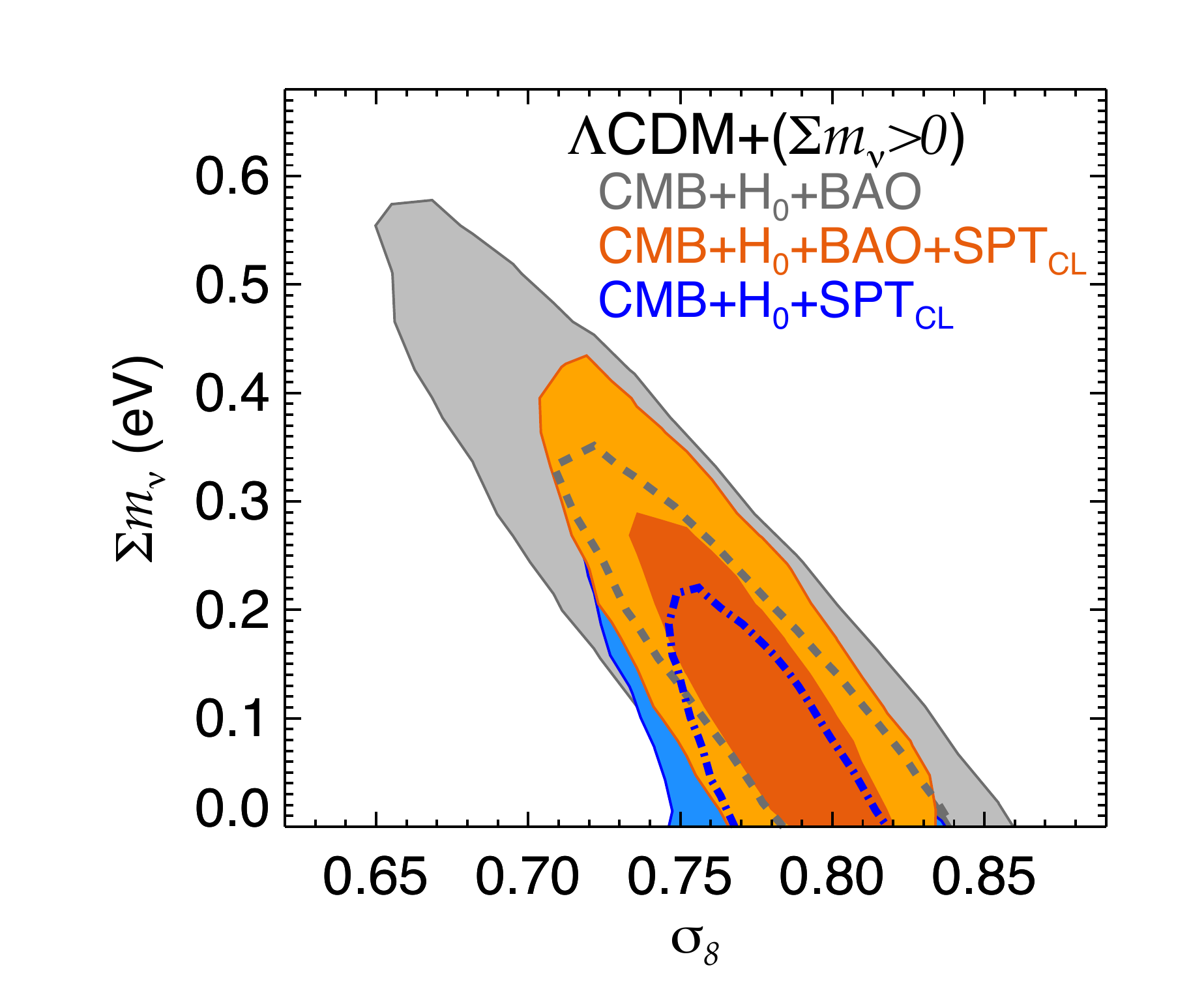}
  \caption[]{Assuming a \LCDM\ cosmology with massive neutrinos, the two-dimensional marginalized 
  constraints on $\Sigma m_{\nu}$ and $\sigma_8$.
  Contours show the 68\% and 95\% confidence regions for the CMB+H$_0$+BAO (gray, dashed), CMB+H$_0$+BAO+\sptcl\ (orange, solid), and 
  CMB+H$_0$+\sptcl\ (blue, dot-dashed) data sets.  The \sptcl\ data improves the constraints on $\sigma_8$ and \sumnu, 
  by factors of 1.8 and 1.4, respectively.
        \\
}
\label{fig:mnu}
\end{figure}

\begin{figure*}[]
\centering
\includegraphics[scale=0.9]{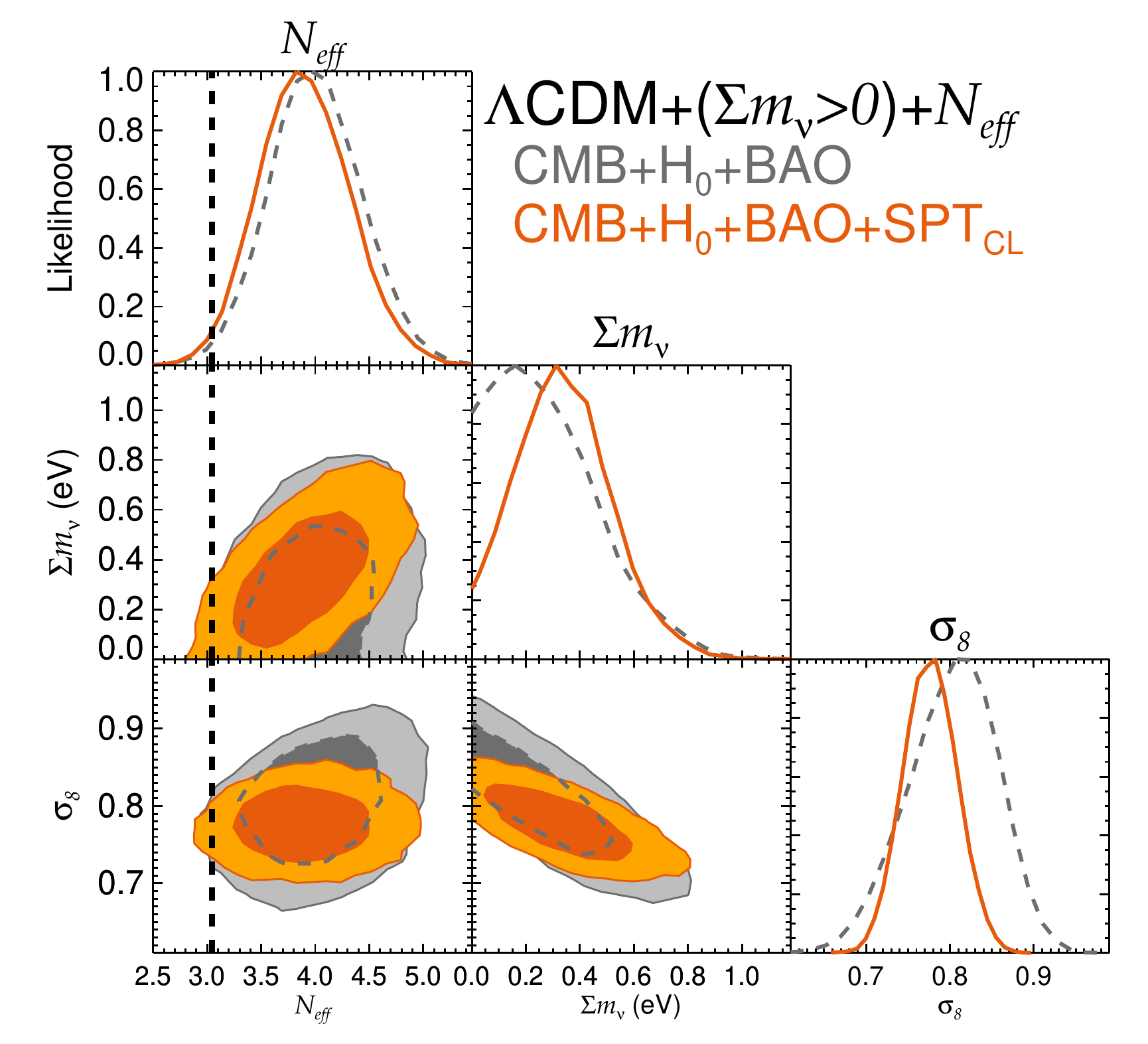}
  \caption[]{Assuming a \LCDM\ cosmology with massive neutrinos and a free number of relativistic species, \neff, the 
  constraints on $\sigma_8$, $\Sigma m_{\nu}$, and \neff.  The plots along the diagonal are the one-dimensional 
  marginalized likelihood.  The off-diagonal plots are the two-dimensional marginalized constraints showing the 
  68\% and 95\% confidence regions.  We show the constraints for the  
  CMB+H$_0$+BAO (gray, dashed), and CMB+H$_0$+BAO+\sptcl\ (orange, solid) data sets.  
  The black vertical line shows \neff$=3.046$, the expected value for three neutrino species.  
   Using the CMB+\hst+BAO+\sptcl\ data set, the 68\% confidence marginalized constraints are $\sigma_8=0.777 \pm 0.031$, \neff$ = 3.91 \pm 0.42$, 
   and \sumnu$=0.34\pm0.17$ eV, with a 95\% CL of \sumnu$ < 0.63$ eV.
   }
\label{fig:neff}
\end{figure*}

\subsection{\LCDM\ with Massive Neutrinos}
\label{sec:mnu}

We next consider a \LCDM\ cosmology with non-zero neutrino masses.  
Cosmological measurements are primarily sensitive to the neutrino masses through their effect 
on structure formation.  A massive neutrino additionally affects the CMB power spectrum 
if it was non-relativistic at the redshift of recombination.  For example, if the heaviest neutrino 
had a mass $\lesssim 0.6$ eV, it would be relativistic at recombination 
and therefore would not significantly affect the structure in the CMB \citep{komatsu09}.  
However, as the universe expanded and cooled neutrinos would transition to 
non-relativistic, and would contribute to $\Omega_m$ but not to structure formation below 
their free streaming scale, implying a lower $\sigma_8$ at $z=0$ and fewer clusters. 
This implies that measurements of the CMB power spectrum alone cannot constrain 
the neutrino mass to significantly less than 0.6 eV per species (i.e., \sumnu $\lesssim 1.8$ eV), and 
the constraints will be significantly degenerate with $\sigma_8$.  
Local measurements of structure break this degeneracy, 
and significantly improve the neutrino mass constraints.  

We also note the significant degeneracy between the CMB power spectrum constraints on \hst\ and \sumnu.  
Massive neutrinos affect the amplitude of the early integrated Sachs-Wolfe effect causing 
a shift of the first peak of the CMB power spectrum towards larger angular scales that can be absorbed by a lower value of
\hst\  \citep{ichikawa05}.
\citet{komatsu11} used a combination of WMAP7+\hst+BAO data to set a limit 
of \sumnu$ < 0.58$ at a 95\% confidence limit (CL).   Following \citet{komatsu11}, we
consider the same combination of data sets to add to the CMB power spectrum measurements, 
which were chosen because of their insensitivity to systematic errors and their 
ability to maximally constrain \sumnu\ by breaking the degeneracy with \hst.
We define \sumnu$= 94$ eV$(f_{\nu} \Omega_c h^2)$, where $f_{\nu}$ is the fraction 
of dark matter in the form of massive neutrinos.  

In Figure \ref{fig:mnu}, we show the constraints on $\sigma_8$ and \sumnu, 
using the CMB+\hst+BAO data set, before and after including the \sptcl\ data. 
In Table \ref{tab:mnu}, we give the marginalized constraints on each parameter.
Using the CMB+\hst+BAO+\sptcl\ data set, we constrain 
\sumnu$ < 0.33$ eV at a 95\% CL, a factor of 1.4 improvement over the constraints without 
the \sptcl\ data.  This improvement is primarily due to the tighter constraints
on $\sigma_8$ for which the uncertainty decreased by a factor 1.8.  
The constraint is lower by excluding the BAO data; using
only the CMB+\hst+\sptcl\ data set we constrain \sumnu$ < 0.28$ eV at a 95\% CL.  These improved 
constraints can be understood from the \hst\ measurements, as also noted in Section \ref{sec:wcdm}.
The results of \citet{riess11} favor a marginally higher \hst\ value than the CMB+BAO data. 
Because of the degeneracy between \sumnu\ and \hst\ in the 
CMB constraints, a higher value of \hst\ tends to favor lower values of \sumnu.  The constraints 
presented here are comparable to other recent results using optically and X-ray selected cluster samples 
with similar cosmological data sets \citep{reid10, mantz10c}.
  
\subsubsection{Number of Relativistic Species}
\label{sec:neff}

Recent measurements have shown a $\sim$2$\sigma$ preference for increased damping in the 
tail of the CMB power spectrum \citep{dunkley11, keisler11}.  This damping could be caused 
by several different physical mechanisms, such as a high primordial helium abundance, 
a running of the scalar spectral index, or additional relativistic species.  This last
explanation is particularly timely because of recent measurements from 
atmospheric \citep{aguilar10} and nuclear reactor \citep{mention11} neutrino oscillation experiments
that find some evidence for a sterile neutrino species.  It has been pointed out that these 
measurements are most consistent with two sterile neutrinos and \sumnu$ \gtrsim 1.7$ eV \citep{kopp11}.  
Therefore, we consider the joint cosmological constraints on \neff\ and \sumnu\ to compare 
with these terrestrial results.  

With only three neutrino species, we would expect \neff$=3.046$, a 
value slightly larger than three because of energy injection from electron-positron 
annihilation at the end of neutrino freeze-out \citep{dicus82, lopez99, mangano05}.  As \neff\ increases,
the contribution to the gravitational potential of the additional neutrino perturbations boosts 
the early growth of dark matter perturbations \citep{bashinsky04}, which also 
increases $\sigma_8$ \citep{hou11}.  As explained in Section \ref{sec:mnu}, 
adding neutrino mass at the levels considered here only affects the low-redshift universe, 
suppressing structure formation, and lowering $\sigma_8$ at $z=0$.  
Therefore, increasing \neff\ will also allow an increasing \sumnu.  
\citet{keisler11} used a combination of CMB+\hst+BAO data to constrain \sumnu$ < 0.69$ eV at a 95\% CL, 
$\sigma_8 = 0.803 \pm 0.056$, and \neff$ = 3.98 \pm 0.43$. 

In Figure \ref{fig:neff}, we show the constraints on \neff, \sumnu, and $\sigma_8$, 
using the CMB+\hst+BAO data set, before and after including the \sptcl\ data. 
In Table, \ref{tab:mnu} we give the marginalized constraints.  When varying \neff\ we 
assume consistency with BBN for our constraints.  Using the 
CMB+\hst+BAO+\sptcl\ data set, we constrain \sumnu$ < 0.63$ eV at a 95\% CL, $\sigma_8=0.777 \pm 0.031$, and 
\neff$ = 3.91 \pm 0.42$.  Relative to \citet{keisler11}, the addition of the \sptcl\ data 
improves the constraints on $\sigma_8$ by a factor of 1.8, and reduces the upper limit on 
\sumnu\ by a factor of 1.1.  However, the addition of the \sptcl\ data does noticeably 
sharpen the peak in the marginalized one-dimensional likelihood for \sumnu, such that the maximum 
likelihood constraint peaks away from zero, \sumnu$ = 0.34 \pm 0.17$ eV.

As noted in \citet{keisler11}, models of the CMB power spectrum that include increased damping 
are favored at the 1.6-1.9$\,\sigma$ level.  However, even if one accepts the need for an extra parameter
to explain the damping, its physical origin is unclear.  Regardless, considering the \neff\ model 
extension is instructive to help understand the model dependency of the neutrino mass constraints.  
\citet{keisler11} considered three models to explain the excess damping and found that 
the \neff\ model had the most significant effect on $\sigma_8$.  The inclusion of \neff\ also weakens the 
constraints on \sumnu, because of the degeneracies between \neff, $\sigma_8$, and \sumnu.  
In the combined cosmological data set, the \sptcl\ data mainly constrains $\sigma_8$, which helps to 
break this degeneracy and indirectly improve the neutrino mass constraints.  Therefore, 
the \sumnu\ constraint from the \neff\ model can be considered a conservative 
upper limit on \sumnu\ regardless of the physical mechanism for the increased damping.   

\subsection{\LCDM\ with \fnl}

\begin{figure}[]
\centering
\includegraphics[scale=0.47]{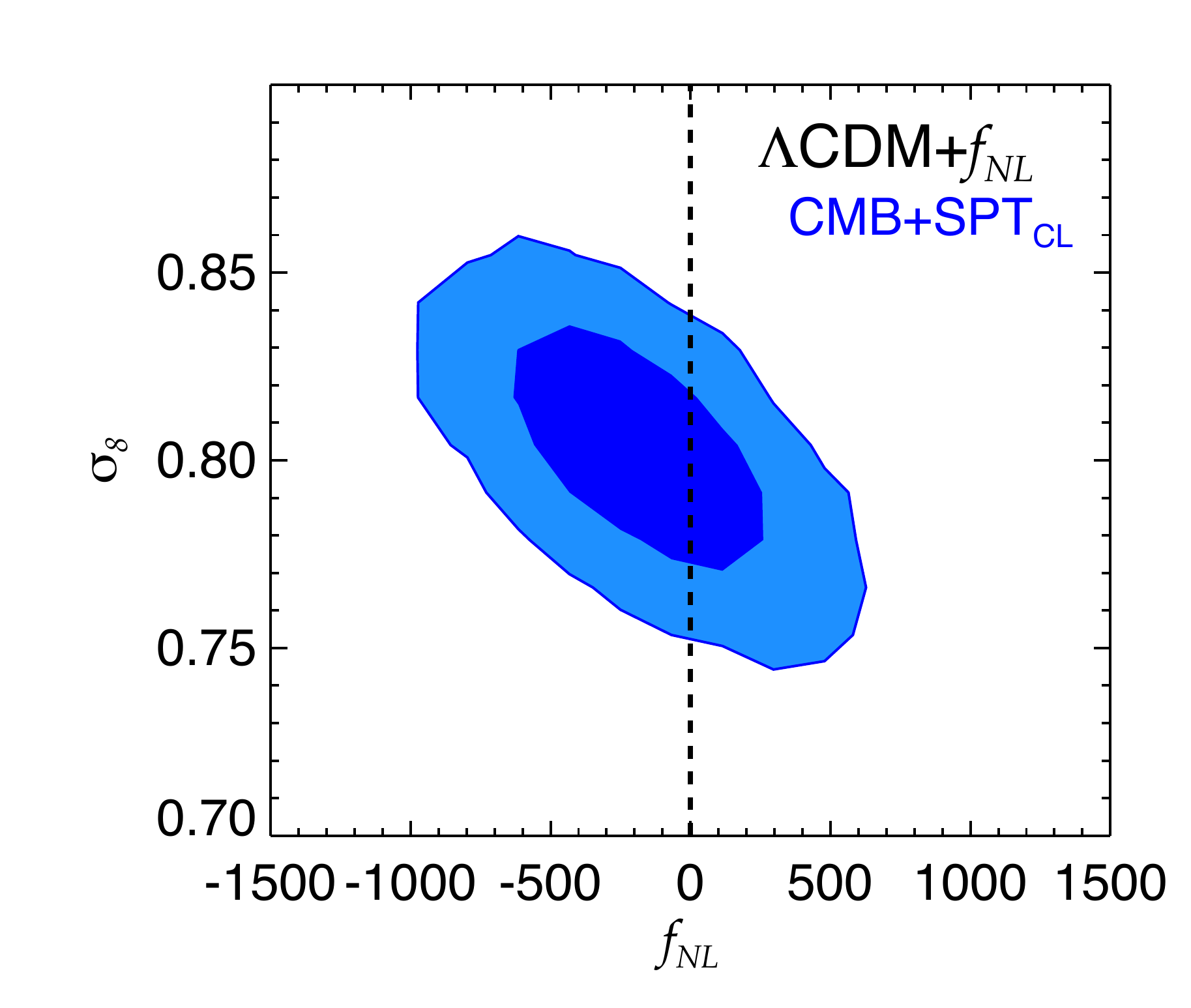}
  \caption[]{Assuming a \LCDM\ cosmology with a primordial non-Gaussianity characterized 
  by the parameter \fnl, the two-dimensional marginalized 
  constraints on $\sigma_8$ and \fnl using the CMB+\sptcl\ data set (blue).   
   Contours show the 68\% and 95\% confidence regions.   
   We only consider the affect of \fnl\ on the \sptcl\ data set.  We measure \fnlspt, consistent with zero non-Gaussianity.
\\
}
\label{fig:fnl}
\end{figure}

Finally, we consider a \LCDM\ cosmology with primordial non-Gaussianity.  
Standard inflationary cosmology predicts that density fluctuations in the universe were 
seeded by random Gaussian fluctuations.  However, inflationary models can be 
constructed that predict significant levels of non-Gaussianity \citep[e.g.,][]{bartolo04}.  
The leading order non-Gaussian term is typically described by the parameter \fnl.  
Using a measurement of the CMB power spectrum from WMAP7 data, 
\citet{komatsu11} measured a 95\% CL of $-10 < \fnl < 74$.  
Primordial non-Gaussianity can also manifest itself through the abundance of massive 
galaxy clusters.  While the constraints from an SPT-like SZ survey are not expected
to be competitive with current CMB constraints \citep{dalal08}, they constitute 
a constraint independent from the CMB results which is sensitive to very different 
physical scales.  In principle, even a single massive high-redshift cluster can falsify 
standard \LCDM\ cosmology \citep{mortonson11}, though currently the most massive 
cluster known at $z > 1$ is not in significant tension \citep{foley11}.  
In our analysis, we incorporate \fnl\ as a modification of the cluster 
mass function following the prescription of \citet{dalal08}, and only consider the effect of \fnl\ 
on the \sptcl\ data set.  

In Figure \ref{fig:fnl}, we show the constraints on $\sigma_8$ and \fnl\ using the CMB+\sptcl\
data set.  The marginalized constraints are \fnlspt, $\sigma_8=0.803 \pm 0.022$, 
and $A_{SZ}=5.27\pm0.89$.  The best-fit value of \fnl\ is slightly negative, generally implying 
fewer massive high-redshift clusters.   In Section \ref{sec:lcdm}, we found that for a \LCDM\ cosmology using the CMB+\sptcl\ 
data, our constraint for $A_{SZ}$ was 0.9$\, \sigma$ lower than the simulation prior.  This 
corresponds to a higher mass for a cluster of a given \SN, which can also be thought of 
as the SPT survey having fewer clusters than expected given the simulation prior.  When \fnl\ 
is added as a parameter, $A_{SZ}$ moves back towards its simulation 
prior, and the deficit of clusters can be maintained by a more negative \fnl.  For the model 
to match the number of clusters measured by SPT, a more negative \fnl\ can be balanced with either  a 
larger $\sigma_8$ or $A_{SZ}$, creating a degeneracy in this direction.  Regardless, any deficit of clusters is not significant 
relative to the uncertainty on either $\sigma_8$ or $A_{SZ}$, even in a \LCDM\ cosmology 
for which they are best constrained.  

Our results are consistent with \citet{williamson11} which used the 26 most massive clusters in the full
2500 \sqdeg\ SPT-SZ survey to constrain \fnl$=20\pm450$.  Our work differs from \citet{williamson11} in 
that we use a much smaller area of the SPT-SZ survey, we select clusters down to a 
lower \SN\ threshold, and we use an improved mass calibration.   

\begin{table*}[thbp]
\begin{center}
\caption[Error Budget]{Error Budget} \footnotesize
\begin{tabular}{l | r | r r | r  r}
\hline \hline
&  \multicolumn{1}{c|}{\LCDM} & \multicolumn{2}{|c|}{\wCDM} &  \multicolumn{2}{|c}{\wCDM} \\
 &  \multicolumn{1}{c|}{\sptcl+\hst+BBN} & \multicolumn{2}{|c|}{\sptcl+\hst+BBN} &  \multicolumn{2}{|c}{CMB+BAO+SNe+\sptcl} \\
 & $\sigma_8 (\Omega_m/0.25)^{0.30}$ & $w$ & $\sigma_8$ & $w$ & $\sigma_8$ \\
 \hline
Baseline, Section \ref{sec:lcdm},\ref{sec:wcdm} &   $0.785\pm0.037$ & $-1.09\pm0.36$ & $0.773\pm0.088$ & $-0.973\pm0.063$ & $0.793\pm0.028$ \\
SNe Systematic &  $-$ & $-$ & $-$ &$\pm0.026$ & $\pm0.005$\\
SZ-\Yx\ Scaling &  $\pm0.010$ &$\pm0.19$ & $\pm0.066$  &$\pm0.013$ & $\pm0.013$ \\
X-ray Scaling Systematic &  $\pm0.028$ &$\pm0.15$ & $\pm0.036$ & $\pm0.019$ & $\pm0.014$ \\
Statistical  & $\pm0.023$ & $\pm0.27$ & $\pm0.046$ & $\pm0.033$ & $\pm0.013$ \\
  \hline \hline
\end{tabular}
\label{tab:syst}
\tablecomments{We give the mean and 68\% confidence intervals for a subset of the 
cosmological parameters reported in Sections \ref{sec:lcdm} and \ref{sec:wcdm}.  The last four rows give 
the 1$\sigma$ error in each cosmological parameter due to the stated uncertainty.  
}
\end{center}
\end{table*}

\section{Sources of Uncertainty}
\label{sec:err}

Previous SPT cluster survey results, namely V10 and  \citet{williamson11}, found  
cosmological constraints that were limited most significantly by the cluster mass 
calibration, or equivalently the fractional uncertainty in $A_{SZ}$.  
In this work, we have reduced this uncertainty  by incorporating 
the external mass calibration from the \yxm\ relation using X-ray observations of the 
SPT clusters.  We can directly estimate the impact of the uncertainties in 
the X-ray and SZ scaling relations by importance sampling the MCMC chains, where we
post-process the chains by imposing a narrow prior on each scaling relation parameter centered 
around the best-fit value.  The resulting increase in precision on the cosmological 
parameters allows a measure of the impact from the uncertainty in the scaling relations.
In this way, we effectively ``fix" the X-ray and scaling relation parameters, 
a process which we will implicitly be referring to throughout this section.  
For a \wCDM\ cosmology, we also consider the impact of the
SNe systematic uncertainty on the cosmological results presented here.  

With enough SZ and X-ray observations, we expect the \snm\ 
calibration to be limited by the calibration of the \yxm\ relation because the latter has tighter external priors.  
In practice, there will be an additional uncertainty in the \snm\ calibration 
from the limited number of SZ and X-ray observations for cross-calibration, 
and this uncertainty will also degrade the cosmological constraints.  
We wish to separate this effect, which we will refer to as the SZ-\Yx\ scaling 
uncertainty, from the additional systematic uncertainty from the 
\yxm\ calibration, which we will refer to as the X-ray scaling uncertainty, and 
the statistical uncertainty from the cluster sample size.  
By fixing the X-ray and SZ scaling relation parameters, as described above,
we can measure the impact of the SZ-\Yx\ scaling uncertainty, X-ray scaling 
uncertainty, and statistical uncertainty on our cosmological constraints.  

\subsection{\LCDM\ Cosmology: Scaling Relation Uncertainty}
\label{sec:lerr}

We first consider the \LCDM\ constraints using the \sptcl+\hst+BBN data set, the results of which 
were described in Section \ref{sec:lcdm}.  This data best 
constrained the combination of $\sigma_8 (\Omega_m/0.25)^{0.30} = 0.785 \pm 0.037$.  
The sources of uncertainty for this result are summarized in Table \ref{tab:syst}, 
and are discussed below.  

For the X-ray scaling relation parameters, only the uncertainty in $A_X$ and $C_X$, 
the normalization and redshift evolution parameters, contribute significantly to the uncertainty 
on $\sigma_8 (\Omega_m/0.25)^{0.30}$.  Fixing each parameter separately 
implies that they contribute an uncertainty on $\sigma_8 (\Omega_m/0.25)^{0.30}$ 
of $\pm$0.022 and $\pm0.015$, respectively.  It is not surprising that the normalization of the mass calibration significantly affects the 
constraints, and the redshift evolution can be understood for similar reasons.  For a cluster at the median 
redshift of the SPT sample, $z=0.74$, the prior on the $C_X$ value effectively contributes an 
additional 7\% to the cluster mass calibration.  This can be compared to the 10\% mass calibration uncertainty from the 
prior on $A_X$. Fixing all X-ray parameters simultaneously, 
implies that they contribute an uncertainty on $\sigma_8 (\Omega_m/0.25)^{0.30}$ of $\pm0.028$.

For the SZ scaling relation parameters, only the uncertainty in $A_{SZ}$ contributes 
significantly to the uncertainty on $\sigma_8 (\Omega_m/0.25)^{0.30}$.  
Fixing all the SZ scaling parameters, we measure an 
uncertainty on $\sigma_8 (\Omega_m/0.25)^{0.30}$ of $\pm0.023$ from statistical 
uncertainty, $\pm0.010$ from the SZ-\Yx\ scaling uncertainty, and $\pm0.028$
due to X-ray scaling uncertainty (as discussed above).  The relatively low contribution 
from the SZ-\Yx\ scaling uncertainty is not surprising 
considering the constraints on the fractional uncertainty of $A_{SZ}$, which was near the 
systematic limit of 14\% imposed by the \yxm\ calibration.

Therefore, the \LCDM\ constraints are nearly at the systematic limit 
from the calibration of the \yxm\ relation.  For our constraint of $\sigma_8 (\Omega_m/0.25)^{0.30} = 0.785 \pm 0.037$, 
the X-ray scaling and statistical uncertainty contribute almost equal amounts of
$\pm0.028$ and $\pm0.023$, respectively.  By only increasing the cluster sample size we could 
reduce the uncertainty by up to $\sim1.3$ ($\sim0.037/0.028$).  Further improvements would require a
more accurate cluster mass calibration.

\subsection{\wCDM\ Cosmology: Scaling Relation and SNe Uncertainty}
\label{sec:werr}

We next consider the sources of uncertainty for the \wCDM\ cosmology discussed in Section \ref{sec:wcdm}.  
We will concentrate on using the \sptcl+\hst+BBN and CMB+BAO+SNe+\sptcl\
data sets, which produce constraints of \wsptwcdm\  and \wallwcdm, respectively.  The sources of uncertainty for this result are 
summarized in Table \ref{tab:syst}, and are discussed below.  

We first consider the \sptcl+\hst+BBN data set.   For the X-ray scaling relation parameters, 
we again find that the uncertainty in $A_X$ and $C_X$ contribute the largest uncertainty on $w$.
Fixing each parameter independently implies they contribute an
uncertainty of $\delta w = \pm 0.10$ and $\pm0.11$, respectively, and a total 
uncertainty of $\pm0.15$. 
For the SZ scaling relation parameters, only the uncertainty in $A_{SZ}$ contributes 
significantly to the uncertainty on $w$.  Fixing all the SZ scaling parameters, 
we measure an uncertainty on $w$ of: $\pm0.27$ from statistical 
uncertainty, $\pm0.19$ from the SZ-\Yx\ scaling uncertainty, and $\pm0.15$
due to X-ray scaling uncertainty.  Therefore, unlike the \LCDM\ case, we find that our constraints on $w$ 
would be significantly improved by adding more clusters and additional \Yx\ measurements.
A similar conclusion is reached repeating the above analysis for $\sigma_8$.  
In principle, adding more clusters and \Yx\ measurements would reduce the uncertainty 
on $w$ and $\sigma_8$ to values limited by the X-ray scaling uncertainty.  
In this limit we should measure $w$ and $\sigma_8$ with an unceratinty of 
$\delta w = \pm 0.15$ and $\delta \sigma_8 = \pm 0.036$, or $\sim2.5$ times better than 
our current constraints.  

When considering the CMB+BAO+SNe+\sptcl\ data set, we reach qualitatively similar 
conclusions, however the total uncertainty is significantly lower because of 
the parameter degeneracies that are broken from the additional data sets.  
We first re-run the MCMC chains without SNe systematic uncertainty.  Fixing 
all the SZ scaling parameters, we measure a 
statistical uncertainty of $\delta w = \pm 0.033$, a factor of two improvement relative to 
the constraints including all systematic uncertainties.  Comparing this uncertainty 
to that with no fixed parameters, we estimate an uncertainty on $w$ of $ \pm0.019$ from X-ray scaling 
uncertainty and $\pm0.013$ from SZ-\Yx\ scaling uncertainty. 
The addition of the \sptcl\ data also significantly reduces the systematic uncertainty from
SNe.  Running a CMB+BAO+SNe MCMC chain with and without SNe systematics, we 
measure $w=-1.014 \pm 0.078$ and $w=-1.017 \pm 0.050$, respectively.  This implies that 
SNe systematics are contributing an uncertainty of $\delta w = \pm 0.060$.  After adding the 
\sptcl\ data the uncertainty from SNe systematics is reduced to  $\delta w = \pm 0.026$, a factor of  $\sim2.3$ improvement. 

\subsection{Point Source Contamination}

In V10, it was argued that point source contamination contributed a negligible 
level of uncertainty relative to the statistical precision of the cluster sample.  
Since we are using the same cluster sample, we expect the same conclusion to hold, though 
we briefly summarize their arguments here.  From Poisson distributed sources, the probability of a chance 
superposition of a bright point source ($\gtrsim 6$ mJy) with a cluster is negligible, given 
the sky density of sources at 150 GHz ($\sim 1$ deg$^{-2}$, \citet{vieira10}).
Furthermore, our cosmological analysis in Section \ref{sec:cosmology} explicitly accounts for a Poisson 
distributed background of sources, and the X-ray measurements 
gives an additional systematic check on an offset in the SZ measurements.  
Correlated emission from cluster members could potentially fill in cluster decrements. 
However, correlated radio emission has previously been calculated to be negligible at 150 GHz for clusters of 
the typical SPT mass scale and redshift range \citep{lin09, sehgal10}.   In V10, it was  
also argued that the level of correlated dusty emission is negligible,
from the known quenching of star formation in massive clusters \citep{hashimoto98},
and the sub-millimeter luminosity function \citep{pascale09}.  In addition, more recent 
\spitzer\ infrared observations of a sample of X-ray selected groups and low-mass clusters, 
found that correlated dusty emission is insignificant compared to the SZ signal \citep{george11}.
These arguments apply to the clusters in the SPT sample, which span a similar redshift range and are of 
higher mass than the objects considered in that work.  

\section{Discussion}
\label{sec:disc}

\subsection{Improvement Relative to V10}

In this work, the cluster sample is the same as used in V10 for their 
cosmological analysis.  However, we have improved the cosmological constraints relative to V10
by including X-ray measurements in order to reduce the cluster mass calibration uncertainty.   
It is not straightforward to quantify the improvement
for two main reasons.  First, V10 used pre-existing WMAP7 MCMC chains 
from \citet{komatsu11} that they importance sampled by re-weighting the chains by 
the likelihood of the SPT cluster catalog given each set of parameters.  In this work, we generated 
new MCMC chains while simultaneously fitting both data sets. Second, each result
uses somewhat different external data sets, in particular for the CMB power spectrum 
measurements, where the results in this work also include CMB measurements from \citet{keisler11}.  

Without explicitly correcting for these differences, we can approximate the improvement  from 
including the X-ray measurements by considering the relative improvements of adding the \sptcl\ 
data to the CMB data used in either result.  For a \LCDM\ cosmology, the WMAP7 
data constrains $\sigma_8=0.801 \pm 0.030$.  In V10, the addition of the SPT data, 
this constraint improved to $\sigma_8=0.791 \pm 0.027$, a factor of 1.1 improvement.  
In this work, for a  \LCDM\ cosmology using the \sptcl\ and CMB data, we 
constrained \sigalllcdm, a factor of 1.5 improvement over the constraints from the CMB alone.  
Therefore, the addition of the X-ray measurements improved the \LCDM\ constraints on $\sigma_8$ by a factor of $\sim$1.4.  
A comparison of the \wCDM\ cosmological constraints 
is more complicated because of the somewhat different handling of the external data sets.  Regardless, 
the significant improvement in the constraints from the X-ray measurements is clear.  

\subsection{Prospects for Further Improvement}

The results in this paper were derived using 18 clusters from 178 \sqdeg\ of the 2500 \sqdeg\ SPT-SZ survey. 
The full survey will significantly increasing both the area and 
overall depth of the SZ maps.   
\citet[][in prep.]{reichardt12} will present a catalog of $\sim$200 clusters from the first  800 \sqdeg\, of the SPT-SZ survey, 
with a median redshift of $\sim0.5$ and a median mass of $\mass \sim  2.3 \times 10^{14} M_{\odot} / h$.
This sample is representative of the cluster yield and properties for the full survey, which was completed in November 2011
and has detected $\sim$500 clusters.  
Therefore it is useful to consider how the method 
used in this work will be applied to the full survey, and what level of improvement we can expect on the
cosmological constraints.   

Using the \sptcl+\hst+BBN data set, we found that our constraints are currently limited by both statistical 
uncertainty and the SZ-\Yx\ scaling uncertainty.
Both uncertainties would be improved by adding more SPT clusters with additional \Yx\ measurements.  
Recently the SPT collaboration was awarded a \chandra\ X-ray Visionary Project (XVP) to 
complete X-ray observations of the 80 
most significant clusters at $z > 0.4$ detected in the first 2000 \sqdeg\ of the SPT-SZ survey.  
As argued in Section \ref{sec:wcdmsnm}, we would need $\gtrsim 50$ clusters with \Yx\ measurements for the 
\snm\ calibration to be limited by the \yxm\ uncertainty.  With this many clusters, the statistical uncertainty 
on $w$ should decrease to a level below the systematic uncertainty from the 
X-ray scaling relation, $\delta w = \pm 0.15$.  Combining the full 2500 \sqdeg\ SPT-SZ survey 
with the \chandra\ XVP observations, we would be limited to this constraint from the 
current calibration of the \yxm\ relation. 

To reduce the systematic uncertainty further, we would need more accurate cluster mass estimates
than currently exist from X-ray measurements alone.  
In Section \ref{sec:werr}, we found that the X-ray scaling systematics were currently limited
by the uncertainty in $A_X$ and $C_X$, whose fractional uncertainty was 10\% and 50\%, respectively.  
Reducing their uncertainty by a factor of two, would reduce their contribution to the systematic 
uncertainty to $\delta w = \pm 0.037$ and $\pm0.074$, respectively.   
This would effectively correspond to an overall mass calibration uncertainty of 5\% with an 
additional 6\% uncertainty in the evolution of the mass calibration between $z=0.0 - 1.1$.  

This level of mass calibration should be achievable by incorporating additional data sets, such as optical 
velocity dispersion \citep{white10} or weak lensing measurements \citep{hoekstra07, becker11}.  
For example, in massive clusters, the scatter in weak lensing mass estimates is expected to be $\sim$20\% \citep{becker11}.  
Therefore, with weak lensing observations of two sets of $\sim$15-20 clusters at low and high-redshift, 
this level of accuracy should be achievable.  Towards this goal, the SPT collaboration has been approved for 
weak lensing observations of $\sim$35 SPT-detected clusters spanning $0.30 < z < 1.3$ using the 
Magellan and \hubble\ telescopes.  Additionally, the SPT collaboration has been approved for optical velocity 
dispersion observations of $\sim$100 SPT-detected clusters using the Very Large Telescope (VLT) 
and a large NOAO program on Gemini South.  With these data sets, we expect to achieve the factor of 
two improvement in mass calibration, as discussed above.  

Applying this calibration to the full 2500 \sqdeg\ \sptcl+\hst+BBN data set, we should 
constrain $w$ with an accuracy of $\sim$8\%, or a factor of $\sim$4.5 tighter than the current \sptcl+\hst+BBN constraints.
This improved constraint would be comparable to the current constraints from the 
CMB+BAO+SNe data, and would be an independent systematic test of the 
standard dark energy paradigm by measuring the effect of dark energy on the 
growth of structure.   Combining the existing CMB+BAO+SNe data with the 
2500 \sqdeg\ SPT cluster sample, the uncertainty from the SZ-\Yx\ scaling and the cluster sample size
is expected to be negligible compared to the uncertainty contributed by the improved cluster mass 
calibration.  In this case, the SPT cluster data would contribute an uncertainty of only $\sim 1\%$
to the significantly improved constraint on $w$ from the combined data set.  
 
\section{Conclusions}
\label{sec:con}

We use measurements from the SPT-SZ cluster survey in combination with X-ray measurements 
to constrain cosmological parameters.  We have described and implemented 
a method that simultaneously fits for cosmological parameters and the scaling 
of the SZ and X-ray observables with cluster mass.   
The method is generalizable to multiple cluster observables, and 
self-consistently accounts for the effects of cluster selection and uncertainties in cluster mass calibration 
on the derived cosmological constraints.  We apply this method to a 
SZ-selected catalog of 18 galaxy clusters identified in 178 \sqdeg\ of the 2500 \sqdeg\ 
SPT-SZ survey.  This is the first analysis of an SZ survey 
to directly incorporate X-ray observations, which has reduced the uncertainty on 
both the cluster mass calibration and the cosmological constraints.  

For a \LCDM\ cosmology, we find that the \sptcl+\hst+BBN data best constrain 
$\sigma_8 (\Omega_m/0.25)^{0.30} = 0.785 \pm 0.037$, where the total uncertainty 
consists of an approximately equal amount of statistical and systematic uncertainty.  These constraints are 
consistent, and comparable, with other constraints using X-ray-selected \citep{vikhlinin09,mantz10b} 
and optically-selected \citep{rozo10} cluster samples.  In combination with 
measurements of the CMB power spectrum from the SPT data and the 
seven-year WMAP data, the SPT cluster data constrain \sigalllcdm\ and \omalllcdm, a factor of 1.5
improvement on each parameter over the constraints from the CMB data alone.  

We consider several extensions beyond a \LCDM\ cosmological model by including the 
following as free parameters: the dark energy equation of state ($w$), the sum of the neutrino masses (\sumnu), the effective number of 
relativistic species (\neff), and a primordial non-Gaussianity ($f_{NL}$).  

For a \wCDM\ cosmology, the \sptcl+\hst+BBN data constrain \wsptwcdm\ and \sigsptwcdm, consistent 
with dark energy being due to a cosmological constant, and with comparable uncertainties
to constraints from the CMB data alone.  Using the CMB+BAO+SNe+\sptcl\ data set, 
we constrain \wallwcdm\ and \sigallwcdm, 
a factor of 1.25 and 1.4 improvement, respectively, over the constraints without 
SPT cluster data.  The uncertainty on $w$ consists of approximately equal contributions 
from statistical uncertainty, systematic uncertainty from SNe, and systematic uncertainty 
from cluster scaling relations, with the latter contributing an uncertainty of $\delta w = \pm 0.023$.  

We next consider a \LCDM\ cosmology with a non-zero neutrino mass.  Using a 
CMB+\hst+BAO+\sptcl\ data set, we constrain the sum of the neutrino masses \sumnu\ to be $<0.33$ eV at 
95\% confidence, a factor of 1.4 improvement over the constraints 
without SPT cluster data.  We find even tighter constraints when we exclude the 
BAO data set, which tend to favor a lower value of \hst\ and therefore a higher neutrino mass.  
Using a CMB+\hst+\sptcl\ data set, we constrain \sumnu$<0.28$ eV
at 95\% confidence.   We also consider a model with a free effective number of 
relativistic species, \neff, to explain the increased damping that is observed in the 
the CMB power spectrum.  Using a CMB+\hst+BAO+\sptcl\ data set, we jointly 
measure \neff$=3.91 \pm 0.42$ and \sumnu$=0.34\pm0.17$ eV, while constraining \sumnu$ < 0.63$ eV at 95\% confidence.  

Finally, we consider a \LCDM\ cosmology where we allow the number of observed 
clusters to be affected by non-Gaussian density fluctuations 
characterized by the parameter \fnl.  Using a CMB+\sptcl\ data set, we measure \fnlspt, 
consistent with zero non-Gaussianity. 

The results presented in this paper use 18 clusters from 178 \sqdeg\ of the 2500 \sqdeg\ SPT-SZ survey, 
and are limited by the combination of the cluster sample size and mass calibration. 
The SPT-SZ survey was completed in November 2011, and has detected $\sim$500 clusters
with a median redshift of $\sim0.5$ and a median mass of $\mass \sim  2.3 \times 10^{14} M_{\odot} / h$.
Ongoing X-ray, weak lensing, and optical velocity dispersion observations of 
SPT-SZ-selected clusters will be used to produce an improved cluster mass calibration of the sample.  
The full SPT-SZ survey and improved mass calibration will produce 
constraints on $w$ comparable to current constraints from the combination of 
CMB+BAO+SNe data, and would represent an independent systematic test of the 
standard dark energy paradigm by measuring the effect of dark energy on the 
growth of structure.  The combination of CMB+BAO+SNe data with the 
SPT cluster sample will break degeneracies between the data sets resulting 
in significantly tighter constraints on dark energy.   


\acknowledgments

The South Pole Telescope program is supported by the National Science
Foundation through grant ANT-0638937.  Partial support is also
provided by the NSF Physics Frontier Center grant PHY-0114422 to the
Kavli Institute of Cosmological Physics at the University of Chicago,
the Kavli Foundation, and the Gordon and Betty Moore Foundation.  
Additional data were obtained with the 6.5~m
Magellan Telescopes located at the Las Campanas Observatory,
Chile. Support for X-ray analysis was provided by NASA through Chandra
Award Numbers 12800071, 12800088, and G02-13006A issued by the Chandra X-ray
Observatory Center, which is operated by the Smithsonian Astrophysical
Observatory for and on behalf of NASA under contract NAS8-03060.
Optical imaging data from the Blanco 4~m at Cerro Tololo Interamerican
Observatories (programs 2005B-0043, 2009B-0400, 2010A-0441,
2010B-0598) and spectroscopic observations from VLT programs
086.A-0741 and 286.A-5021 and Gemini program GS-2009B-Q-16 were
included in this work.
We acknowledge the use of the Legacy Archive for
Microwave Background Data Analysis (LAMBDA).  Support for LAMBDA is
provided by the NASA Office of Space Science.  Galaxy cluster research
at Harvard is supported by NSF grant AST-1009012.  Galaxy cluster
research at SAO is supported in part by NSF grants AST-1009649 and
MRI-0723073.  The McGill group acknowledges funding from the National
Sciences and Engineering Research Council of Canada, Canada Research
Chairs program, and the Canadian Institute for Advanced Research.
X-ray research at the CfA is supported through NASA Contract NAS
8-03060.  This work is based in part on observations made with the Spitzer Space Telescope, which is operated by the Jet Propulsion Laboratory, California Institute of Technology under a contract with NASA. Support for this work was provided by NASA through an award issued by JPL/Caltech. The Munich group acknowledges support from the Excellence
Cluster Universe and the DFG research program TR33.  R.J.F.\ is
supported by a Clay Fellowship.  B.A.B\ is supported by a KICP
Fellowship, M.Bautz acknowledges support from contract
2834-MIT-SAO-4018 from the Pennsylvania State University to the
Massachusetts Institute of Technology.  M.D.\ acknowledges support
from an Alfred P.\ Sloan Research Fellowship, W.F.\ and C.J.\
acknowledge support from the Smithsonian Institution, and B.S.\
acknowledges support from the Brinson Foundation.

{\it Facilities:}
\facility{Blanco (MOSAIC)},
\facility{CXO (ACIS)},
\facility{Gemini-S (GMOS)},
\facility{Magellan:Baade (IMACS)},
\facility{Magellan:Clay (LDSS3)},
\facility{Spitzer (IRAC)},
\facility{South Pole Telescope},
\facility{XMM-Newton (EPIC)}

\bibliography{../../BIBTEX/spt}

\newpage
\appendix

\section{X-ray Observations and Results}
\label{ref:appxray}

In Table \ref{tab:xrayobs}, we give updated X-ray observables for the clusters used in this work, 
as discussed in Section \ref{sec:xray}.
For the five clusters without new measurements, we give the results from A11 directly, 
in order to provide a complete listing for the cluster sample.  In Table \ref{tab:xrayids}, 
we give the complete list of \chandra\ observation identifications (ObsIDs) used for 
clusters with new \chandra\ observations  

\begin{table}[h]
\begin{minipage}{\textwidth}
\centering
\caption{X-ray Observables for SPT clusters} \small
\begin{tabular}{lrrrrr}
\hline\hline
\rule[-2mm]{0mm}{6mm}
Name  & $z$ & \rclust & \Tx & \Mg & \Yx    \\
 & & (kpc) & (keV) & ($10^{13}$ $M_\sun$) & ($10^{14}$ $M_\sun$ keV) \\
\hline
SPT-CL J0509-5342\tablenotemark{a}  & 0.463 & $1062 \pm 39$ & $7.0^{+1.4}_{-1.1}$ & $5.6^{+0.2}_{-0.2}$ & $4.3 \pm 0.8$ \\
SPT-CL J0528-5300\tablenotemark{b}  & 0.765 & $765 \pm 47$ & $5.2^{+1.9}_{-1.2}$ & $2.8^{+0.3}_{-0.3}$ & $1.6 \pm 0.5$ \\
SPT-CL J0533-5005\tablenotemark{b}  & 0.881 & $666 \pm 51$ & $3.9^{+1.6}_{-1.1}$ & $2.3^{+0.5}_{-0.4}$ & $1.0 \pm 0.4$ \\
SPT-CL J0546-5345\tablenotemark{b}  & 1.067 & $823 \pm 27$ & $6.8^{+1.2}_{-0.9}$ & $7.4^{+0.4}_{-0.3}$ & $4.8 \pm 0.8$ \\
SPT-CL J0551-5709\tablenotemark{b}  & 0.423 & $923 \pm 34$ & $4.0^{+0.6}_{-0.6}$ & $5.1^{+0.6}_{-0.6}$ & $1.9 \pm 0.4$ \\
SPT-CL J0559-5249\tablenotemark{a}  & 0.611 & $1071 \pm 30$ & $7.7^{+1.1}_{-0.8}$ & $8.3^{+0.3}_{-0.2}$ & $6.4 \pm 0.8$ \\
SPT-CL J2331-5051\tablenotemark{a}  & 0.571 & $972 \pm 34$ & $5.9^{+1.3}_{-0.8}$ & $5.7^{+0.2}_{-0.2}$ & $3.5 \pm 0.6$ \\
SPT-CL J2332-5358\tablenotemark{c}  & 0.403 & $1166 \pm 31$ & $7.8^{+1.0}_{-0.9}$ & $7.6^{+0.2}_{-0.3}$ & $6.1 \pm 0.8$ \\
SPT-CL J2337-5942\tablenotemark{a}  & 0.781 & $1046 \pm 39$ & $8.9^{+2.0}_{-1.4}$ & $9.5^{+0.4}_{-0.6}$ & $8.5 \pm 1.7$ \\
SPT-CL J2341-5119\tablenotemark{a}  & 0.998 & $847 \pm 37$ & $8.0^{+1.9}_{-1.6}$ & $5.6^{+0.2}_{-0.2}$ & $4.7 \pm 1.0$ \\
SPT-CL J2342-5411\tablenotemark{c}  & 1.074 & $648 \pm 29$ & $5.0^{+0.9}_{-0.8}$ & $2.6^{+0.3}_{-0.3}$ & $1.4 \pm 0.3$ \\
SPT-CL J2355-5056\tablenotemark{c}  & 0.320 & $997 \pm 31$ & $5.3^{+0.9}_{-0.7}$ & $3.9^{+0.2}_{-0.1}$ & $2.2 \pm 0.4$ \\
SPT-CL J2359-5009\tablenotemark{b,c}  & 0.774 & $778 \pm 36$ & $5.2^{+1.3}_{-0.9}$ & $3.1^{+0.3}_{-0.3}$ & $1.8 \pm 0.4$ \\
SPT-CL J0000-5748\tablenotemark{c}  & 0.701 & $950 \pm 68$ & $8.3^{+3.6}_{-2.2}$ & $4.4^{+0.5}_{-0.5}$ & $4.2 \pm 1.6$ \\
\hline
\end{tabular}
\label{tab:xrayobs}
\begin{@twocolumnfalse}
\tablecomments{X-ray observables for clusters with \chandra\ or \xmm\ observations.  
For clusters with new spectroscopic redshifts or new \chandra\ observations, we have 
recalculated their X-ray observables, as described in Section \ref{sec:xray}.  To maintain 
consistency with A11, all X-ray observables are calculated assuming a preferred \LCDM\ cosmology 
using {\sl WMAP}7+BAO+{\sl $H_0$} data with $\Omega_M = 0.272$, $\Omega_\Lambda = 0.728$ and
$H_0 = 70.2~$km$~$s$^{-1}~$Mpc$^{-1}$ \citep{komatsu11}.
}
\tablenotetext{a}{X-ray observables taken from A11.}
\tablenotetext{b}{Updated for new \chandra\ observations.}
\tablenotetext{c}{Updated for new spectroscopic redshift.}
\end{@twocolumnfalse}
\normalsize
\end{minipage}
\end{table}

\begin{table}[h]
\begin{minipage}{\textwidth}
\centering
\caption{Clusters with new \chandra\ X-ray Observations} \small
\begin{tabular}{lr|rr|rr|l}
\hline\hline
\rule[-2mm]{0mm}{6mm}
 & & \multicolumn{2}{|c|}{A11} & \multicolumn{2}{|c|}{This Work} & \\
\hline
Name    &  $z$     &  Exposure   & Source & Exposure & Source  & ObsIDs     \\
 & & [ks] & Counts & [ks] & Counts & \\
\hline
SPT-CL J0528-5300 & 0.765 & 36.5 & 356 & 115.9 & 1732 & 9341, 10862, 11996, {\bf 11747, 11874, 12092, 13126} \\
SPT-CL J0533-5005 & 0.881 & 41.5 & 201 & 67.7 & 344 & 11748, 12001, {\bf 12002} \\
SPT-CL J0546-5345 & 1.067 & 55.6 & 1304 & 67.8 & 1512 & 9332, 9336, 10851, 10864, {\bf 11739}  \\
SPT-CL J0551-5709 & 0.423 & 19.8 & 876 & 33.2 & 1212 & 11871, {\bf 11743} \\
SPT-CL J2359-5009 & 0.774 & 57.9 & 713 & 122.4 & 1522 & 9334, 11742, 11864, {\bf 11997} \\
\hline
\end{tabular}
\label{tab:xrayids}
\begin{@twocolumnfalse}
\tablecomments{The ObsIDs refer to all the \chandra\ observations used in this work.  The ObsIDs that 
are new, relative to A11, are highlighted in {\bf bold}.  For clusters not listed here, we use the same
\chandra\ and \xmm\ observations as listed by A11. 
}
\end{@twocolumnfalse}
\normalsize
\end{minipage}
\end{table}


\section{Likelihood modification to account for cosmological dependence of \Yx}
\label{app:yx_term}

In Section \ref{sec:like} we presented a procedure for translating the theoretical mass function, ${dN}/{dM dz}$, into 
observable space,  ${dN}/{d\SN d\Yx dz}$.
Under the assumption that this transformation is independent of the cosmological and scaling relation parameters $\vec{p}$ we are ultimately trying to recover, 
this procedure modifies the log likelihood by a constant offset.

However, in the case of \Yx, this assumption ceases to hold true as \Yx\ is a derived quantity, calculated explicitly for each new value of $\vec{p}$ (i.e., at each likelihood evaluation in the MCMC). This added subtlety can be addressed in the following fashion: Let us define \Yxstar\ as the value of \Yx\ when evaluated at some reference point in parameter space $\vec{p}^*$. In order for the probability contained in a differential volume to be independent of a change of variables, we need to multiply by the Jacobian of the transformation, as follows
\begin{equation}
  P(\vec{z},\vec{\xi},\vec{Y}_x^* | \vec{p}) = \left| \frac{\partial(\vec{Y}_x)}{\partial(\vec{Y}_x^*)} \right| P(\vec{z},\vec{\xi},\vec{Y}_x | \vec{p}) = \left| \prod_i f_i(\vec{p}) \right| P(\vec{z},\vec{\xi},\vec{Y}_x | \vec{p})
\end{equation}
where $f_i(\vec{p})$ is the ratio of the \Yx\ value of the $i^\mathrm{th}$ cluster at $\vec{p}$ to its value at $\vec{p}^*$. Expressing this in terms of log probability, and ignoring constant offsets, we obtain
\begin{equation}
  \ln P(\vec{z},\vec{\xi},\vec{Y}_x | \vec{p}) = \ln P(\vec{z},\vec{\xi},\vec{Y}_x^* | \vec{p}) + \sum_i \ln Y_{x,i}
\end{equation}
This results in the straightforward prescription of adding $\sum{\ln{\Yx}}$ to the likelihood at each step in the MCMC, a process very similar to that  suggested and performed in \citet{mantz08} and \citet{vikhlinin09b}.

\section{Mass Estimates}
\label{app:mass_ests}

We present posterior mass estimates for all 18 clusters considered in this work in Table \ref{tab:mass_ests}. Where applicable, these are joint X-ray and SZ posterior mass estimates, for clusters without X-ray data we use the SZ posterior mass estimate.  We calculate a probability density function on a mass grid at each point in the \LCDM\ chain that was calculated using the CMB+\sptcl\ data, from Section \ref{sec:lcdmcosmo}.  The probability density functions are combined to obtain a mass estimate that has been fully marginalized over all cosmological and scaling relation parameters.  We report the mean and the 68\% confidence interval for the mass estimate.  

\begin{table*}[h]
\begin{minipage}{\textwidth}
  \centering
  \caption{Mass Estimates for the SPT Cluster Catalog} \small
  \begin{tabular}{l c c c}
    \hline \hline
    \rule[-2mm]{0mm}{6mm}
    Object Name	& $\SN$ & $z$ & $M_{500}(\rho_{crit})(10^{14}\,\msun h_{70}^{-1})$ \\
    \hline
SPT-CL J0509-5342 &  6.61 &  0.463 & 5.11  $\pm$   0.68  \\
SPT-CL J0511-5154\tablenotemark{a} &  5.63 &  0.74 & 3.36  $\pm$   0.86  \\
SPT-CL J0521-5104\tablenotemark{a} &  5.45 &  0.72 & 3.21  $\pm$   0.86  \\
SPT-CL J0528-5259 &  5.45 &  0.765 & 2.96  $\pm$   0.54  \\
SPT-CL J0533-5005 &  5.59 &  0.881 & 2.54  $\pm$   0.54  \\
SPT-CL J0539-5744\tablenotemark{a} &  5.12 &  0.77 & 2.93  $\pm$   0.86  \\
SPT-CL J0546-5345 &  7.69 &  1.067 & 4.79  $\pm$   0.64  \\
SPT-CL J0551-5709 &  6.13 &  0.423 & 3.61  $\pm$   0.54  \\
SPT-CL J0559-5249 &  9.28 &  0.611 & 6.36  $\pm$   0.79  \\
SPT-CL J2301-5546\tablenotemark{a} &  5.19 &  0.748 & 3.00  $\pm$   0.86  \\
SPT-CL J2331-5051 &  8.04 &  0.572 & 4.89  $\pm$   0.68  \\
SPT-CL J2332-5358 &  7.30 &  0.403 & 6.21  $\pm$   0.79  \\
SPT-CL J2337-5942 & 14.94 &  0.781 & 7.68  $\pm$   1.04  \\
SPT-CL J2341-5119 &  9.65 &  0.998 & 5.14  $\pm$   0.71  \\
SPT-CL J2342-5411 &  6.18 &  1.074 & 2.75  $\pm$   0.46  \\
SPT-CL J2355-5056 &  5.89 &  0.320 & 3.96  $\pm$   0.54  \\
SPT-CL J2359-5009 &  6.35 &  0.774 & 3.32  $\pm$   0.54  \\
SPT-CL J0000-5748 &  5.48 &  0.701 & 4.04  $\pm$   0.68  \\
    \hline
  \end{tabular}
  \label{tab:mass_ests}
  \tablenotetext{a}{These clusters have only SZ data, and no X-ray observations.}
  \normalsize
\end{minipage}
\end{table*}

\end{document}